\documentclass[lettersize,journal]{IEEEtran}

\usepackage{amsmath,amsfonts,amssymb}
\usepackage{algorithmic}
\usepackage{algorithm}
\usepackage{array}
\usepackage[caption=false,font=normalsize,labelfont=sf,textfont=sf]{subfig}
\usepackage{textcomp}
\usepackage{stfloats}
\usepackage{soul}
\sethlcolor{green}
\usepackage{url}
\usepackage{verbatim}
\usepackage{graphicx}
\usepackage{tikz}
\usetikzlibrary{patterns}
\usepackage{pgfplots}
\usepgfplotslibrary{groupplots}
\pgfplotsset{compat=1.14}
\usepackage{cite}
\usepackage{comment}
\usepackage{booktabs}
\usepackage{multirow}
\usepackage{microtype}
\usepackage{balance}
\usepackage{enumitem}
\usepackage[draft]{hyperref}

\usepackage[acronym,nonumberlist,nopostdot,nogroupskip,automake=false,
            nomain]{glossaries}
\newignoredglossary{hidden}
\makenoidxglossaries
\newacronym{3gpp}{3GPP}{3rd Generation Partnership Project}
\newacronym{4g}{4G}{4th generation}
\newacronym{5g}{5G}{5th generation}
\newacronym{6g}{6G}{6th generation}
\newacronym{5gc}{5GC}{5G Core}
\newacronym{adc}{ADC}{Analog to Digital Converter}
\newacronym{aerpaw}{AERPAW}{Aerial Experimentation and Research Platform for Advanced Wireless}
\newacronym{ai}{AI}{Artificial Intelligence}
\newacronym{aimd}{AIMD}{Additive Increase Multiplicative Decrease}
\newacronym{am}{AM}{Acknowledged Mode}
\newacronym{amc}{AMC}{Adaptive Modulation and Coding}
\newacronym{amf}{AMF}{Access and Mobility Management Function}
\newacronym{aops}{AOPS}{Adaptive Order Prediction Scheduling}
\newacronym{api}{API}{Application Programming Interface}
\newacronym{apn}{APN}{Access Point Name}
\newacronym{ap}{AP}{Application Protocol}
\newacronym{aqm}{AQM}{Active Queue Management}
\newacronym{ausf}{AUSF}{Authentication Server Function}
\newacronym{avc}{AVC}{Advanced Video Coding}
\newacronym{awgn}{AGWN}{Additive White Gaussian Noise}
\newacronym{arfcn}{ARFCN}{Absolute Radio Frequency Channel Number}
\newacronym{balia}{BALIA}{Balanced Link Adaptation Algorithm}
\newacronym{bbu}{BBU}{Base Band Unit}
\newacronym{bdp}{BDP}{Bandwidth-Delay Product}
\newacronym{ber}{BER}{Bit Error Rate}
\newacronym{bf}{BF}{Beamforming}
\newacronym{bler}{BLER}{Block Error Rate}
\newacronym{brr}{BRR}{Bayesian Ridge Regressor}
\newacronym{bs}{BS}{Base Station}
\newacronym{bsr}{BSR}{Buffer Status Report}
\newacronym{bss}{BSS}{Business Support System}
\newacronym{ca}{CA}{Carrier Aggregation}
\newacronym{caas}{CaaS}{Connectivity-as-a-Service}
\newacronym{cb}{CB}{Code Block}
\newacronym{cc}{CC}{Congestion Control}
\newacronym{ccid}{CCID}{Congestion Control ID}
\newacronym{cco}{CC}{Carrier Component}
\newacronym{cdd}{CDD}{Cyclic Delay Diversity}
\newacronym{cdf}{CDF}{Cumulative Distribution Function}
\newacronym{cdn}{CDN}{Content Distribution Network}
\newacronym{cn}{CN}{Core Network}
\newacronym{codel}{CoDel}{Controlled Delay Management}
\newacronym{comac}{COMAC}{Converged Multi-Access and Core}
\newacronym{cord}{CORD}{Central Office Re-architected as a Datacenter}
\newacronym{cornet}{CORNET}{COgnitive Radio NETwork}
\newacronym{cosmos}{COSMOS}{Cloud Enhanced Open Software Defined Mobile Wireless Testbed for City-Scale Deployment}
\newacronym{cots}{COTS}{Commercial Off-the-Shelf}
\newacronym{cp}{CP}{Control Plane}
\newacronym{cyp}{CP}{Cyclic Prefix}
\newacronym{up}{UP}{User Plane}
\newacronym{cpu}{CPU}{Central Processing Unit}
\newacronym{cqi}{CQI}{Channel Quality Information}
\newacronym{cr}{CR}{Cognitive Radio}
\newacronym{cran}{CRAN}{Cloud \gls{ran}}
\newacronym{crs}{CRS}{Cell Reference Signal}
\newacronym{csi}{CSI}{Channel State Information}
\newacronym{csirs}{CSI-RS}{Channel State Information - Reference Signal}
\newacronym{cu}{CU}{Central Unit}
\newacronym{cho}{CHO}{Conditional Handover}
\newacronym{d2tcp}{D$^2$TCP}{Deadline-aware Data center TCP}
\newacronym{d3}{D$^3$}{Deadline-Driven Delivery}
\newacronym{dac}{DAC}{Digital to Analog Converter}
\newacronym{dag}{DAG}{Directed Acyclic Graph}
\newacronym{das}{DAS}{Distributed Antenna System}
\newacronym{dash}{DASH}{Dynamic Adaptive Streaming over HTTP}
\newacronym{dc}{DC}{Dual Connectivity}
\newacronym{dccp}{DCCP}{Datagram Congestion Control Protocol}
\newacronym{dce}{DCE}{Direct Code Execution}
\newacronym{dci}{DCI}{Downlink Control Information}
\newacronym{dctcp}{DCTCP}{Data Center TCP}
\newacronym{dl}{DL}{Deep Learning}
\newacronym{dmr}{DMR}{Deadline Miss Ratio}
\newacronym{dmrs}{DMRS}{DeModulation Reference Signal}
\newacronym{drlcc}{DRL-CC}{Deep Reinforcement Learning Congestion Control}
\newacronym{drs}{DRS}{Discovery Reference Signal}
\newacronym{du}{DU}{Distributed Unit}
\newacronym{e2e}{E2E}{end-to-end}
\newacronym{ecaas}{ECaaS}{Edge-Cloud-as-a-Service}
\newacronym{ecn}{ECN}{Explicit Congestion Notification}
\newacronym{edf}{EDF}{Earliest Deadline First}
\newacronym{embb}{eMBB}{Enhanced Mobile Broadband}
\newacronym{empower}{EMPOWER}{EMpowering transatlantic PlatfOrms for advanced WirEless Research}
\newacronym{enb}{eNB}{evolved Node Base}
\newacronym{endc}{EN-DC}{E-UTRAN-\gls{nr} \gls{dc}}
\newacronym{epc}{EPC}{Evolved Packet Core}
\newacronym{eps}{EPS}{Evolved Packet System}
\newacronym{es}{ES}{Edge Server}
\newacronym{etsi}{ETSI}{European Telecommunications Standards Institute}
\newacronym[firstplural=Estimated Times of Arrival (ETAs)]{eta}{ETA}{Estimated Time of Arrival}
\newacronym{eutran}{E-UTRAN}{Evolved Universal Terrestrial Access Network}
\newacronym{faas}{FaaS}{Function-as-a-Service}
\newacronym{fapi}{FAPI}{Functional Application Platform Interface}
\newacronym{fdd}{FDD}{Frequency Division Duplexing}
\newacronym{fdm}{FDM}{Frequency Division Multiplexing}
\newacronym{fdma}{FDMA}{Frequency Division Multiple Access}
\newacronym{fed4fire}{FED4FIRE+}{Federation 4 Future Internet Research and Experimentation Plus}
\newacronym{fir}{FIR}{Finite Impulse Response}
\newacronym{fit}{FIT}{Future \acrlong{iot}}
\newacronym{fpga}{FPGA}{Field Programmable Gate Array}
\newacronym{fr2}{FR2}{Frequency Range 2}
\newacronym{fs}{FS}{Fast Switching}
\newacronym{fscc}{FSCC}{Flow Sharing Congestion Control}
\newacronym{ftp}{FTP}{File Transfer Protocol}
\newacronym{fw}{FW}{Flow Window}
\newacronym{ge}{GE}{Gaussian Elimination}
\newacronym{gnb}{gNB}{Next Generation Node Base}
\newacronym{gop}{GOP}{Group of Pictures}
\newacronym{gpr}{GPR}{Gaussian Process Regressor}
\newacronym{gpu}{GPU}{Graphics Processing Unit}
\newacronym{gtp}{GTP}{GPRS Tunneling Protocol}
\newacronym{gtpc}{GTP-C}{GPRS Tunnelling Protocol Control Plane}
\newacronym{gtpu}{GTP-U}{GPRS Tunnelling Protocol User Plane}
\newacronym{gtpv2c}{GTPv2-C}{\gls{gtp} v2 - Control}
\newacronym{gw}{GW}{Gateway}
\newacronym{harq}{HARQ}{Hybrid Automatic Repeat reQuest}
\newacronym{hetnet}{HetNet}{Heterogeneous Network}
\newacronym{hh}{HH}{Hard Handover}
\newacronym{hol}{HOL}{Head-of-Line}
\newacronym{hqf}{HQF}{Highest-quality-first}
\newacronym{hss}{HSS}{Home Subscription Server}
\newacronym{http}{HTTP}{HyperText Transfer Protocol}
\newacronym{ia}{IA}{Initial Access}
\newacronym{iab}{IAB}{Integrated Access and Backhaul}
\newacronym{ic}{IC}{Incident Command}
\newacronym{ietf}{IETF}{Internet Engineering Task Force}
\newacronym{ims}{IMS}{Infrastructure Management Service}
\newacronym{imsi}{IMSI}{International Mobile Subscriber Identity}
\newacronym{imt}{IMT}{International Mobile Telecommunication}
\newacronym{iot}{IoT}{Internet of Things}
\newacronym{ip}{IP}{Internet Protocol}
\newacronym{itu}{ITU}{International Telecommunication Union}
\newacronym{kpi}{KPI}{Key Performance Indicator}
\newacronym{kpm}{KPM}{Key Performance Measurement}
\newacronym{kvm}{KVM}{Kernel-based Virtual Machine}
\newacronym{los}{LOS}{Line-of-Sight}
\newacronym{lsm}{LSM}{Link-to-System Mapping}
\newacronym{lstm}{LSTM}{Long Short Term Memory}
\newacronym{lte}{LTE}{Long Term Evolution}
\newacronym{lxc}{LXC}{Linux Container}
\newacronym{m2m}{M2M}{Machine to Machine}
\newacronym{mac}{MAC}{Medium Access Control}
\newacronym{manet}{MANET}{Mobile Ad Hoc Network}
\newacronym{mano}{MANO}{Management and Orchestration}
\newacronym{mc}{MC}{Multi-Connectivity}
\newacronym{mcc}{MCC}{Mobile Cloud Computing}
\newacronym{mchem}{MCHEM}{Massive Channel Emulator}
\newacronym{mcs}{MCS}{Modulation and Coding Scheme}
\newacronym{mec}{MEC}{Multi-access Edge Computing}
\newacronym{mec2}{MEC}{Mobile Edge Cloud}
\newacronym{mfc}{MFC}{Mobile Fog Computing}
\newacronym{mgen}{MGEN}{Multi-Generator}
\newacronym{mi}{MI}{Mutual Information}
\newacronym{mib}{MIB}{Master Information Block}
\newacronym{miesm}{MIESM}{Mutual Information Based Effective SINR}
\newacronym{mimo}{MIMO}{Multiple Input, Multiple Output}
\newacronym{ml}{ML}{Machine Learning}
\newacronym{mlr}{MLR}{Maximum-local-rate}
\newacronym[plural=\gls{mme}s,firstplural=Mobility Management Entities (MMEs)]{mme}{MME}{Mobility Management Entity}
\newacronym{mmtc}{mMTC}{Massive Machine-Type Communications}
\newacronym{mmwave}{mmWave}{millimeter wave}
\newacronym{mpdccp}{MP-DCCP}{Multipath Datagram Congestion Control Protocol}
\newacronym{mptcp}{MPTCP}{Multipath TCP}
\newacronym{mr}{MR}{Maximum Rate}
\newacronym{mrdc}{MR-DC}{Multi \gls{rat} \gls{dc}}
\newacronym{mse}{MSE}{Mean Square Error}
\newacronym{mss}{MSS}{Maximum Segment Size}
\newacronym{mt}{MT}{Mobile Termination}
\newacronym{mtd}{MTD}{Machine-Type Device}
\newacronym{mtu}{MTU}{Maximum Transmission Unit}
\newacronym{mumimo}{MU-MIMO}{Multi-user \gls{mimo}}
\newacronym{mvno}{MVNO}{Mobile Virtual Network Operator}
\newacronym{nalu}{NALU}{Network Abstraction Layer Unit}
\newacronym{nas}{NAS}{Non-Access Stratum}
\newacronym{nbiot}{NB-IoT}{Narrow Band IoT}
\newacronym{nfv}{NFV}{Network Function Virtualization}
\newacronym{nfvi}{NFVI}{Network Function Virtualization Infrastructure}
\newacronym{ni}{NI}{Network Interfaces}
\newacronym{nic}{NIC}{Network Interface Card}
\newacronym{nlos}{NLOS}{Non-Line-of-Sight}
\newacronym{now}{NOW}{Non Overlapping Window}
\newacronym{nsm}{NSM}{Network Service Mesh}
\newacronym[type=hidden]{nr}{NR}{New Radio}
\newacronym{nrf}{NRF}{Network Repository Function}
\newacronym{nsa}{NSA}{Non Stand Alone}
\newacronym{nse}{NSE}{Network Slicing Engine}
\newacronym{nssf}{NSSF}{Network Slice Selection Function}
\newacronym{o2i}{O2I}{Outdoor to Indoor}
\newacronym{oai}{OAI}{OpenAirInterface}
\newacronym{oaicn}{OAI-CN}{\gls{oai} \acrlong{cn}}
\newacronym{oairan}{OAI-RAN}{\acrlong{oai} \acrlong{ran}}
\newacronym{oam}{OAM}{Operations, Administration and Maintenance}
\newacronym{ofdm}{OFDM}{Orthogonal Frequency Division Multiplexing}
\newacronym{olia}{OLIA}{Opportunistic Linked Increase Algorithm}
\newacronym{omec}{OMEC}{Open Mobile Evolved Core}
\newacronym{onap}{ONAP}{Open Network Automation Platform}
\newacronym{onf}{ONF}{Open Networking Foundation}
\newacronym{onos}{ONOS}{Open Networking Operating System}
\newacronym{oom}{OOM}{\gls{onap} Operations Manager}
\newacronym{opnfv}{OPNFV}{Open Platform for \gls{nfv}}
\newacronym[type=hidden]{oran}{O-RAN}{O-RAN}
\newacronym{orbit}{ORBIT}{Open-Access Research Testbed for Next-Generation Wireless Networks}
\newacronym{os}{OS}{Operating System}
\newacronym{oss}{OSS}{Operations Support System}
\newacronym{pa}{PA}{Position-aware}
\newacronym{pase}{PASE}{Prioritization, Arbitration, and Self-adjusting Endpoints}
\newacronym{pawr}{PAWR}{Platforms for Advanced Wireless Research}
\newacronym{pbch}{PBCH}{Physical Broadcast Channel}
\newacronym{pcef}{PCEF}{Policy and Charging Enforcement Function}
\newacronym{pcfich}{PCFICH}{Physical Control Format Indicator Channel}
\newacronym{pcrf}{PCRF}{Policy and Charging Rules Function}
\newacronym{pdcch}{PDCCH}{Physical Downlink Control Channel}
\newacronym{pdcp}{PDCP}{Packet Data Convergence Protocol}
\newacronym{pdsch}{PDSCH}{Physical Downlink Shared Channel}
\newacronym{pdu}{PDU}{Packet Data Unit}
\newacronym{pf}{PF}{Proportional Fair}
\newacronym{pgw}{PGW}{Packet Gateway}
\newacronym{phich}{PHICH}{Physical Hybrid ARQ Indicator Channel}
\newacronym{phy}{PHY}{Physical}
\newacronym{pmch}{PMCH}{Physical Multicast Channel}
\newacronym{pmi}{PMI}{Precoding Matrix Indicators}
\newacronym{powder}{POWDER}{Platform for Open Wireless Data-driven Experimental Research}
\newacronym{ppo}{PPO}{Proximal Policy Optimization}
\newacronym{ppp}{PPP}{Poisson Point Process}
\newacronym{prach}{PRACH}{Physical Random Access Channel}
\newacronym{prb}{PRB}{Physical Resource Block}
\newacronym{psnr}{PSNR}{Peak Signal to Noise Ratio}
\newacronym{pss}{PSS}{Primary Synchronization Signal}
\newacronym{pucch}{PUCCH}{Physical Uplink Control Channel}
\newacronym{pusch}{PUSCH}{Physical Uplink Shared Channel}
\newacronym{qam}{QAM}{Quadrature Amplitude Modulation}
\newacronym{qci}{QCI}{\gls{qos} Class Identifier}
\newacronym{qoe}{QoE}{Quality of Experience}
\newacronym{qos}{QoS}{Quality of Service}
\newacronym{quic}{QUIC}{Quick UDP Internet Connections}
\newacronym{rach}{RACH}{Random Access Channel}
\newacronym{ran}{RAN}{Radio Access Network}
\newacronym[firstplural=Radio Access Technologies (RATs)]{rat}{RAT}{Radio Access Technology}
\newacronym{rcn}{RCN}{Research Coordination Network}
\newacronym{rc}{RC}{RAN Control}
\newacronym{rec}{REC}{Radio Edge Cloud}
\newacronym{red}{RED}{Random Early Detection}
\newacronym{renew}{RENEW}{Reconfigurable Eco-system for Next-generation End-to-end Wireless}
\newacronym{rf}{RF}{Radio Frequency}
\newacronym[longplural=Requests for Comments]{rfc}{RFC}{Request for Comments}
\newacronym{rfr}{RFR}{Random Forest Regressor}
\newacronym{ric}{RIC}{RAN Intelligent Controller}
\newacronym{rlc}{RLC}{Radio Link Control}
\newacronym{rlf}{RLF}{Radio Link Failure}
\newacronym{rlnc}{RLNC}{Random Linear Network Coding}
\newacronym{rmr}{RMR}{RIC Message Router}
\newacronym{rmse}{RMSE}{Root Mean Squared Error}
\newacronym{rnis}{RNIS}{Radio Network Information Service}
\newacronym{rr}{RR}{Round Robin}
\newacronym{rrc}{RRC}{Radio Resource Control}
\newacronym{rrm}{RRM}{Radio Resource Management}
\newacronym{rru}{RRU}{Remote Radio Unit}
\newacronym{rs}{RS}{Remote Server}
\newacronym{rsrp}{RSRP}{Reference Signal Received Power}
\newacronym{rsrq}{RSRQ}{Reference Signal Received Quality}
\newacronym{rss}{RSS}{Received Signal Strength}
\newacronym{rssi}{RSSI}{Received Signal Strength Indicator}
\newacronym{rtt}{RTT}{Round Trip Time}
\newacronym{ru}{RU}{Radio Unit}
\newacronym{rw}{RW}{Receive Window}
\newacronym{rx}{RX}{Receiver}
\newacronym{s1ap}{S1AP}{S1 Application Protocol}
\newacronym{sa}{SA}{standalone}
\newacronym{sack}{SACK}{Selective Acknowledgment}
\newacronym{sap}{SAP}{Service Access Point}
\newacronym{sc2}{SC2}{Spectrum Collaboration Challenge}
\newacronym{scef}{SCEF}{Service Capability Exposure Function}
\newacronym{sch}{SCH}{Secondary Cell Handover}
\newacronym{scoot}{SCOOT}{Split Cycle Offset Optimization Technique}
\newacronym{sctp}{SCTP}{Stream Control Transmission Protocol}
\newacronym{sdap}{SDAP}{Service Data Adaptation Protocol}
\newacronym{sdk}{SDK}{Software Development Kit}
\newacronym{sdm}{SDM}{Space Division Multiplexing}
\newacronym{sdma}{SDMA}{Spatial Division Multiple Access}
\newacronym{sdn}{SDN}{Software-defined Networking}
\newacronym{sdr}{SDR}{Software-defined Radio}
\newacronym{seba}{SEBA}{SDN-Enabled Broadband Access}
\newacronym{sgsn}{SGSN}{Serving GPRS Support Node}
\newacronym{sgw}{SGW}{Service Gateway}
\newacronym{si}{SI}{Study Item}
\newacronym{sib}{SIB}{Secondary Information Block}
\newacronym{sinr}{SINR}{Signal to Interference plus Noise Ratio}
\newacronym{sip}{SIP}{Session Initiation Protocol}
\newacronym{siso}{SISO}{Single Input, Single Output}
\newacronym{sla}{SLA}{Service Level Agreement}
\newacronym{sm}{SM}{Service Model}
\newacronym{smf}{SMF}{Session Management Function}
\newacronym{smo}{SMO}{Service Management and Orchestration}
\newacronym{sms}{SMS}{Short Message Service}
\newacronym{smsgmsc}{SMS-GMSC}{\gls{sms}-Gateway}
\newacronym{snr}{SNR}{Signal-to-Noise-Ratio}
\newacronym{son}{SON}{Self-Organizing Network}
\newacronym{sptcp}{SPTCP}{Single Path TCP}
\newacronym{srb}{SRB}{Service Radio Bearer}
\newacronym{srn}{SRN}{Standard Radio Node}
\newacronym{srs}{SRS}{Sounding Reference Signal}
\newacronym{ss}{SS}{Synchronization Signal}
\newacronym{sss}{SSS}{Secondary Synchronization Signal}
\newacronym{ssb}{SSB}{Synchronization Signal Block}
\newacronym{st}{ST}{Spanning Tree}
\newacronym{svc}{SVC}{Scalable Video Coding}
\newacronym{tb}{TB}{Transport Block}
\newacronym{tcp}{TCP}{Transmission Control Protocol}
\newacronym{tdd}{TDD}{Time Division Duplexing}
\newacronym{tdm}{TDM}{Time Division Multiplexing}
\newacronym{tdma}{TDMA}{Time Division Multiple Access}
\newacronym{tfl}{TfL}{Transport for London}
\newacronym{tfrc}{TFRC}{TCP-Friendly Rate Control}
\newacronym{tft}{TFT}{Traffic Flow Template}
\newacronym{tgen}{TGEN}{Traffic Generator}
\newacronym{tip}{TIP}{Telecom Infra Project}
\newacronym{tm}{TM}{Transparent Mode}
\newacronym{to}{TO}{Telco Operator}
\newacronym{tr}{TR}{Technical Report}
\newacronym{trp}{TRP}{Transmitter Receiver Pair}
\newacronym{ts}{TS}{Technical Specification}
\newacronym{tti}{TTI}{Transmission Time Interval}
\newacronym{ttt}{TTT}{Time-to-Trigger}
\newacronym{tx}{TX}{Transmitter}
\newacronym{uas}{UAS}{Unmanned Aerial System}
\newacronym{uav}{UAV}{Unmanned Aerial Vehicle}
\newacronym{udm}{UDM}{Unified Data Management}
\newacronym{udp}{UDP}{User Datagram Protocol}
\newacronym{udr}{UDR}{Unified Data Repository}
\newacronym{ue}{UE}{User Equipment}
\newacronym{uhd}{UHD}{\gls{usrp} Hardware Driver}
\newacronym{ul}{UL}{Uplink}
\newacronym{um}{UM}{Unacknowledged Mode}
\newacronym{uml}{UML}{Unified Modeling Language}
\newacronym{upa}{UPA}{Uniform Planar Array}
\newacronym{upf}{UPF}{User Plane Function}
\newacronym{urllc}{URLLC}{Ultra Reliable and Low Latency Communications}
\newacronym{usa}{U.S.}{United States}
\newacronym{usim}{USIM}{Universal Subscriber Identity Module}
\newacronym{usrp}{USRP}{Universal Software Radio Peripheral}
\newacronym{utc}{UTC}{Urban Traffic Control}
\newacronym{vim}{VIM}{Virtualization Infrastructure Manager}
\newacronym{vm}{VM}{Virtual Machine}
\newacronym{vnf}{VNF}{Virtual Network Function}
\newacronym{volte}{VoLTE}{Voice over \gls{lte}}
\newacronym{voltha}{VOLTHA}{Virtual OLT HArdware Abstraction}
\newacronym{vr}{VR}{Virtual Reality}
\newacronym{vran}{vRAN}{Virtualized \gls{ran}}
\newacronym{vss}{VSS}{Video Streaming Server}
\newacronym{wbf}{WBF}{Wired Bias Function}
\newacronym{wf}{WF}{Waterfilling}
\newacronym{wg}{WG}{Working Group}
\newacronym{wlan}{WLAN}{Wireless Local Area Network}
\newacronym{osm}{OSM}{Open Source \gls{nfv} Management and Orchestration}
\newacronym{pnf}{PNF}{Physical Network Function}
\newacronym{drl}{DRL}{Deep Reinforcement Learning}
\newacronym{mtc}{MTC}{Machine-type Communications}
\newacronym{osc}{OSC}{O-RAN Software Community}
\newacronym{mns}{MnS}{Management Services}
\newacronym{ves}{VES}{\gls{vnf} Event Stream}
\newacronym{ei}{EI}{Enrichment Information}
\newacronym{fh}{FH}{Fronthaul}
\newacronym{fft}{FFT}{Fast Fourier Transform}
\newacronym{laa}{LAA}{Licensed-Assisted Access}
\newacronym{plfs}{PLFS}{Physical Layer Frequency Signals}
\newacronym{ptp}{PTP}{Precision Time Protocol}
\newacronym{cnn}{CNN}{Convolutional Neural Network}
\newacronym{aoa}{AoA}{Angle of Arrival}
\newacronym{xr}{XR}{Extended Reality}
\newacronym{icc}{ICC}{Intelligence Coordination Controller}
\newacronym{smos}{SMOS}{SMO Services}
\newacronym{focom}{FOCOM}{Federated O-Cloud Orchestration and Management}
\newacronym{nfo}{NFO}{Network Function Orchestration}
\newacronym{dms}{DMS}{Deployment Management Service}
\newacronym{llm}{LLM}{Large Language Model}
\newacronym{isac}{ISAC}{Integrated Sensing and Communications}
\newacronym{mig}{MIG}{Multi-Instance GPU}
\newacronym{dsp}{DSP}{Digital Signal Processing}
\newacronym{itil}{ITIL}{Information Technology Infrastructure Library}
\newacronym{dnn}{DNN}{Deep Neural Networks}
\newacronym{rag}{RAG}{Retrieval-Augmented Generation}
\newacronym{swipt}{SWIPT}{Simultaneous Wireless Information and Power Transfer}
\newacronym{wmmse}{WMMSE}{Weighted Minimum Meansquared-Error}
\newacronym{cad}{CAD}{Computer-Aided Design}
\newacronym{cam}{CAM}{Computer-Aided Manufacturing}
\newacronym{cae}{CAE}{Computer-Aided Engineering}
\newacronym{nrps}{NRPS}{Neural Radio Protocol Stack}
\newacronym{ldpc}{LDPC}{Low-Density Parity-Check}
\newacronym{ibn}{IBN}{Intent-Based Networking}
\newacronym{ota}{OTA}{Over-The-Air}

\usepackage{xcolor}
\usepackage{soul}
\sethlcolor{yellow}
\newcommand{\genesis}{\textsc{Genesis}\xspace}
\newcommand{\synapse}{\textsc{Synapse}\xspace}
\newcommand{\synthesize}{\textsc{Synthesize}\xspace}
\newcommand{\testcap}{\textsc{Test}\xspace}
\newcommand{\harden}{\textsc{Harden}\xspace}
\newcommand{\optimize}{\textsc{Optimize}\xspace}

\newcommand{\discover}{\textsc{Discover}\xspace}
\newcommand{\secureit}{\textsc{Secure}\xspace}

\newcommand{\rfsim}{RFSIM\xspace}

\newcommand{\xfiveg}{X5G\xspace}
\newcommand{\ota}{OTA\xspace}
\newcommand{\testrunner}{\textsc{TestRunner}\xspace}
\newcommand{\specanalyzer}{\textsc{SpecAnalyzer}\xspace}
\newcommand{\codeanalyzer}{\textsc{CodeAnalyzer}\xspace}

\newcommand{\codewriter}{\textsc{CodeWriter}\xspace}
\newcommand{\analyzer}{\textsc{Analyzer}\xspace}
\newcommand{\ingest}{\textsc{Ingest}\xspace}
\newcommand{\specwriter}{\textsc{SpecWriter}\xspace}
\newcommand{\regressionAgent}{\textsc{RegressionAgent}\xspace}

\newcommand{\hypoth}{\textsc{HypothesisEngine}\xspace}

\usepackage{xspace}

\usepackage{listings}
\lstdefinestyle{genesisCode}{
  basicstyle=\ttfamily\footnotesize,
  breaklines=true,
  columns=fullflexible,
  keepspaces=true,
  frame=single,
  captionpos=b,
  showstringspaces=false,
  xleftmargin=4pt,
  xrightmargin=4pt
}
\usepackage{booktabs}
\usepackage{multirow}
\usepackage{pifont}
\usepackage{xcolor}
\usepackage{colortbl}
 
\newcommand{\cmark}{\textcolor{green!50!black}{\ding{51}}}
\newcommand{\xmark}{\textcolor{red!60!black}{\ding{55}}}
\newcommand{\pmark}{\textcolor{orange!85!black}{$\bullet$}}
\definecolor{genesisrow}{RGB}{230, 240, 252}

\definecolor{draft_ready}{RGB}{0, 120, 80}
\definecolor{pending}{RGB}{20, 70, 140}

\begin{document}

%\title{Let There Be RAN: Agentic, Autonomous AI-RAN Synthesis and Testing with \genesis}

% -----------------------
\title{\genesis: Harnessing AI Agents for Autonomous 6G RAN Synthesis, Research, and Testing}

\author{%
  Tamerlan Aghayev, Maxime Elkael, Michele Polese, Minh Dat Nguyen, Gabriele Gemmi, Andrea Lacava,\\Ali Saeizadeh, Reshma Prasad, Paolo Testolina, Angelo Feraudo, Soumendra Nanda,\\Pedram Johari, Salvatore D'Oro, Tommaso Melodia 
  \thanks{The authors are with the Institute for Intelligent Networked Systems at Northeastern University, Boston, MA. Email: melodia@northeastern.edu}%
  \thanks{This work is partially supported by OUSD(R\&E) through Army Research Laboratory Cooperative Agreement Number W911NF-24-2-0065. The views and conclusions contained in this document are those of the authors and should not be interpreted as representing the official policies, either expressed or implied, of the Army Research Laboratory or the U.S.\ Government. The U.S.\ Government is authorized to reproduce and distribute reprints for Government purposes notwithstanding any copyright notation herein. This work is also partially supported by the U.S.\ NSF under award TI-2449452 and under award CNS-2112471.}
}

\maketitle

\glsunset{5g}
\glsunset{6g}

\begin{abstract}

Cellular research and development (R\&D) is throttled by six structural processes that each consume months of manual engineering work per iteration: (i) synthesizing new features from standards or research papers into production code; (ii) conformance and interoperability testing; (iii) hardening against field anomalies and diverse deployment environments; (iv) data-driven optimization of network functionalities; (v) discovering and prototyping novel waveforms, functionalities, and capabilities for future standards; and (vi) securing the stack against vulnerabilities. 
Although \glspl{llm} have compressed comparable R\&D work in general software engineering from days to minutes, their known pitfalls worsen on \gls{ran} use cases: they hallucinate \glspl{api} and mis-read specifications, 
which kills interoperability of \gls{ran} components at the first mistake, and they heavily rely on simulations for designing algorithms, which is notorious for breaking when transferred to real hardware.

To address these challenges, we present \genesis, an agentic \gls{ai} framework that converts intents (e.g., a specification clause, a telemetry anomaly, or a research hypothesis) into solutions validated with over-the-air experiments, fed back into a persistent knowledge base. \genesis is built on three composable primitives (\emph{agents}, \emph{skills}, \emph{hooks}) and a knowledge layer (\synapse) that doubles as the source of ground truth and the recipient of every artifact the framework produces, making capabilities compound across runs. To prototype \genesis, we developed 6 agentic-driven pipelines that synthesize, test, harden, optimize, discover, and secure a network. The design of \genesis uniquely anchors each agentic step in observations and automated tests
that (i) are autonomously executed on heterogeneous cellular
infrastructure and testbeds (from RAN simulators to over-the-air
O-RAN and 5G stacks) and (ii) provide critical feedback to the agents’
decisions. This allows agents to quickly understand errors and shortcomings in their strategies, and converge to implementations that actually work on real-world 5G systems. We present three case studies that together exercise the full R\&D life-cycle: (i) synthesizing and testing the implementation of the 3GPP \texttt{RRC.ConnMean} \gls{kpm}; (ii) synthesizing, testing, and hardening \gls{cho} with a closed-loop xApp over E2SM-RC; and (iii) researching new \gls{ran} scheduling variants, by taking a research hypothesis through code, integration, and comparison with the state of the art. Across multiple statistically independent experiments, \genesis has a 100\% success rate in implementing new stack features, while our baseline (Claude Code with Opus 4.7) consistently fails at each attempt. This paper introduces the \genesis architecture, implementation, and an extensive set of experimental results that profile the harness (e.g., token utilization, cost across different \glspl{llm}) and validate the features that \genesis synthesized. 

\end{abstract}


\glsresetall
\glsunset{3gpp}
\glsunset{nr}

\section{Introduction}
\label{sec:introduction}

Cellular networks research and development is structurally slow. This stems from months-long engineering cycles to bring ideas to products, through research, prototyping, testing, hardening, optimization, evaluation, and standardization. Those stages recur and together form the typical \gls{ran} R\&D life-cycle. 
To address this bottleneck, this paper proposes \genesis, an agentic framework that reduces this effort from months to hours. In this introduction, we discuss the structural bottlenecks and discuss how \genesis addresses these limitations.

\subsection{Challenges in \gls{ran} R\&D}

Six bottlenecks dominate \gls{ran} R\&D: (i)~\emph{synthesizing} new features, from standards clauses or research papers into prototype and production code; (ii)~\emph{testing}, including conformance and interoperability across vendor stacks; (iii)~\emph{hardening} against field anomalies and diverse deployment environments; (iv)~\emph{optimizing} network functionalities through data-driven policies; (v)~\emph{discovering} novel waveforms, functionalities, and capabilities and preparing them for future standards; and (vi)~\emph{securing} the stack while identifying vulnerabilities. Each of these is today a separate engineering arc, burdened with the cost of research, development, and integration with complex multi-vendor systems, \gls{ran} infrastructure, testbeds, and eventually field deployments, even when the underlying intent is small.

While the transition from monolithic, hardware-based appliances to software-driven systems has introduced openness and programmability in the \gls{ran}, it has also exacerbated these issues. A feature today crosses more interfaces, more vendors, and compute platforms than in the appliance/black box era. This breadth, more than any single component, is what shapes the engineering work necessary to carry a feature from a specification clause, or research idea, to a working radio. An analysis of commits and merge requests for the 5G stack in~\cite{openairinterface_gitlab} reveals that the interval between a first code change to a substantial feature merged in the repository's stable branch is 74 days on average, and 207 in the 90th percentile (excluding bug fixes, documentation).

\begin{figure*}
    \centering
    \includegraphics[width=0.9\linewidth]{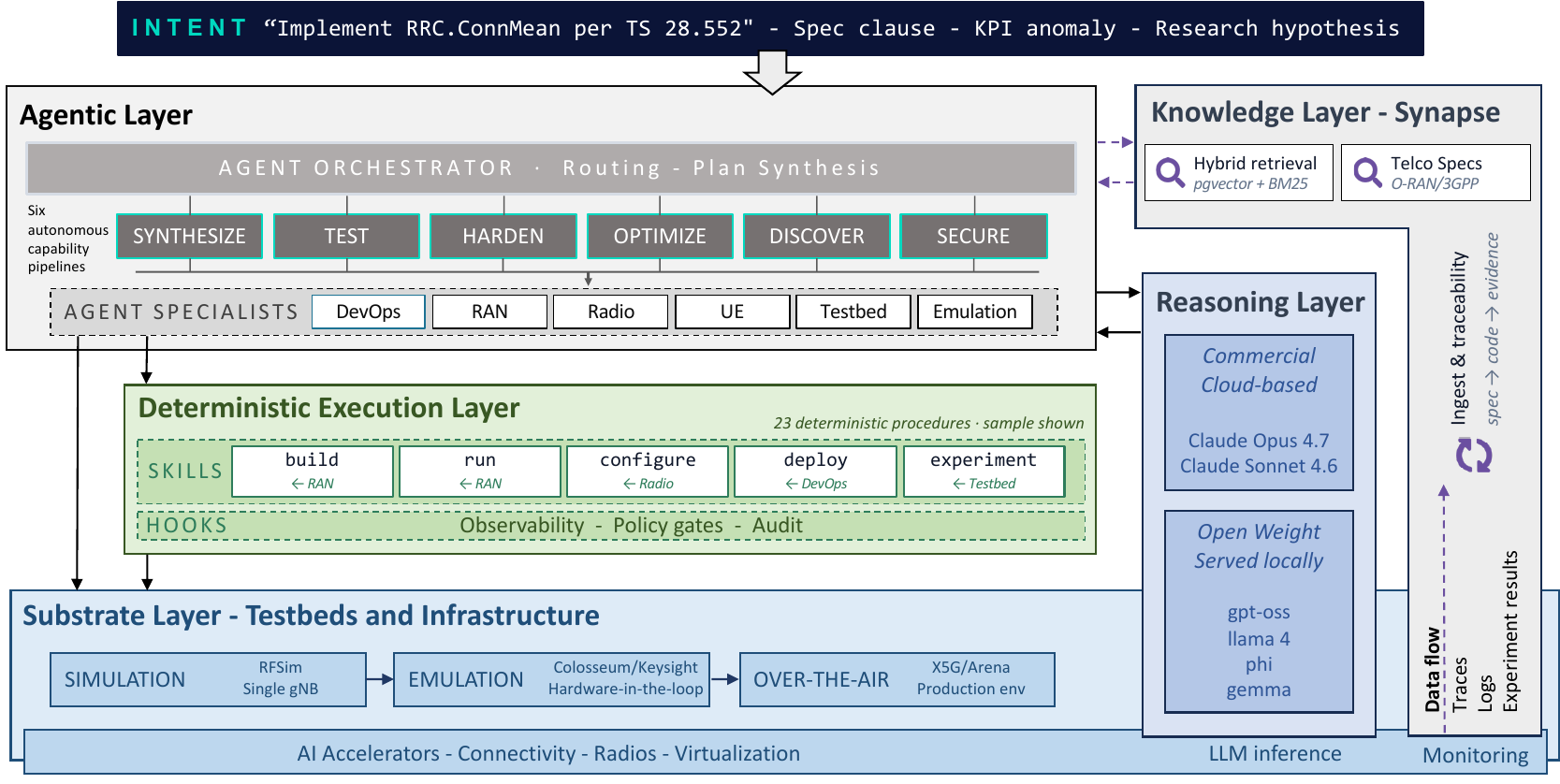}
\caption{\genesis architecture, organized as four horizontal layers tied together by the \synapse knowledge plane and a pluggable \gls{llm} backend. A single intent (a specification clause, a \gls{kpi} anomaly, or a research hypothesis) enters at the top. The \emph{agentic framework} routes it through one of six capability pipelines (\synthesize, \testcap, \harden, \optimize, \discover, \secureit), composed at run-time from a pool of agent specialists (DevOps, RAN, Radio, UE, Testbed, Emulation, among others). The \emph{deterministic execution} layer hosts $\sim$23 parameterized skills (\texttt{build}, \texttt{run}, \texttt{configure}, \texttt{deploy}, \texttt{experiment}, $\ldots$) and hooks for observability, policy gates, and audit. The \emph{substrate layer} provides a three-tier validation continuum from \rfsim through emulation (Colosseum with hardware-in-the-loop) to \ota deployment on production-grade testbeds (X5G, Arena). \synapse serves as both source of ground truth (3GPP/O-RAN specs via hybrid retrieval) and recipient of every artifact each run produces (traces, logs, code diffs with spec-to-code traceability). LLM instances are pluggable, mixing commercial cloud models (e.g., Claude Opus~4.7, Sonnet~4.6) with open-weight models served locally (e.g., gpt-oss, Llama~4, Phi, Gemma). The whole system runs as a closed loop in which the testbed and the knowledge plane co-evolve with the agentic framework.}
    \label{fig:genesis-intro}
\end{figure*}

This complexity delays standardized features from reaching production stacks. Vendors carry most of the engineering cost, and they spend it on what operators are willing to pay for. Network slicing, \gls{urllc}, and \gls{iab} illustrate this: extremely promising in the early 5G era for industrial automation~\cite{sachs2018industrial5g}, disaster recovery~\cite{madapatha2020on}, and similar applications, they are rarely deployed today~\cite{maghsoudnia2024ultra}. These are high-value, low-volume features: each matters in real industrial or defense scenarios, but none is justified by the multi-billion-dollar market that drives traditional telco economics. Openness was supposed to bypass that gate: \glspl{ric} would let operators add control features without changes to the vendor \gls{ran} stack~\cite{polese2023understanding}. In practice, this potential has not materialized: production stacks expose only limited telemetry and control, severely constraining the use-case-tailored AI optimization that the recently formed AI-RAN Alliance~\cite{airan2026} has put on the industry agenda.

\subsection{Software Engineering and \glspl{llm}}

For traditional software development, \gls{llm} agents have the potential to revert this dynamic by lowering engineering lead time and cost. Multi-agent frameworks (AutoGen~\cite{wu2024autogen}, MetaGPT~\cite{hong2023metagpt}, CrewAI~\cite{crewai2026}, Claude Code~\cite{claude-agents-doc}) split work across specialized personas and produce strong results on SWE-bench~\cite{jimenez2023swe}, web automation, and code competitions~\cite{yu2024codereval,daniotti2026who}. However, those wins do not yet transfer to \gls{ran} engineering. As we will show later in this paper, the same agentic frameworks hallucinate \glspl{api}, misinterpret ambiguous specifications, and produce code that may compile, or even run in simulation, but breaks once it attempts at interoperating with other standard devices and interfaces with physical systems such as radios, RF test equipment, commercial \glspl{ue}, and over-the-air testbeds.

Closing this gap is not a pure prompt-engineering problem. Having a closed-loop testing harness is critical: agents may fail, and thus need to be able to test and iterate autonomously on multiple environments with increasing degree of realism. The agents thus need RAN-specific scaffolding: deterministic procedures to, e.g., run experiments, assist in troubleshooting, and operate testbeds, step-by-step pipelines that incorporate verified domain knowledge, and a testbed-based loop that validates every change against real radio behavior. 

\subsection{\genesis}

\genesis is that scaffolding. Figure~\ref{fig:genesis-intro} shows its architecture, with four horizontal layers tied together by a shared knowledge plane and the \gls{llm} backend.

\textbf{Intent layer.} Every run starts from a high-level intent. The figure illustrates three examples of forms the intent can take across the \gls{ran} R\&D life-cycle: a specification clause (\emph{``implement \texttt{RRC.ConnMean} per \gls{ts}~28.552''}), an anomaly on telemetry or \glspl{kpi} observed in production, or a research hypothesis to evaluate. All three enter the framework through the same interface and are routed to the appropriate capability pipeline.

\textbf{Agentic Layer.} Below the intent sits the orchestration plane. An \emph{agent orchestrator} performs routing and defines the plan that maps the intent onto one of six \emph{capability pipelines}. Together, these cover the full R\&D life-cycle: \synthesize (specification-to-code), \testcap (conformance, interoperability, and regression testing), \harden (anomaly-to-fix loops), \optimize (data-driven algorithmic training and deployment), \discover (novel capabilities or features), and \secureit (adversarial analysis of the stack). Each pipeline is assembled at run-time from a pool of \emph{agent specialists} (e.g., DevOps, RAN, Radio, UE, Testbed, Emulation) whose personas encode expertise in a specific domain, decision authority, and the subset of tools they may invoke. Agents \emph{reason}, and leverage the execution layer to take procedural steps.

\textbf{Deterministic Execution Layer.} This layer hosts the \emph{skills}, i.e., a set of two dozen parameterized, deterministic procedures with explicit success criteria (\texttt{build}, \texttt{run}, \texttt{configure}, \texttt{deploy}, \texttt{experiment}, and more). Skills execute based on reasoning provided by the agents. In the same layer, \emph{hooks} are event-driven shell commands that fire around every action and provide three cross-cutting planes: observability (structured event logs that bypass the agent's context window), policy gates (non-bypassable safety checks on actions that touch critical components, e.g., radios), and audit (provenance records for every change). The agent/skill/hook split is what makes \genesis composable, observable, and portable across agentic runtimes.

\textbf{Substrate Layer: the Agentic Infrastructure.} The same agents and skills that write the code also drive the infrastructure on which the code runs. Three tiers form a \emph{validation continuum}: pure simulation (\rfsim with a single \gls{gnb}); emulation (Colosseum~\cite{bonati2021colosseum}, Keysight instruments, or other hardware-in-the-loop emulators); and \ota deployment on production-grade testbeds (X5G~\cite{villa2025x5g}, Arena~\cite{bertizzolo2020arena}). The continuum spans a vast compute infrastructure, including AI accelerators, fronthaul connectivity, real radios, and a virtualization fabric. Every test outcome at every tier is fed back to the agent's next decision, so an autonomously generated change is not merely written but exercised under increasingly realistic conditions. The agentic testbed leverages automation and virtualization capabilities based on our AutoRAN framework~\cite{maxenti2025autoranautomatedzerotouchopen}. With this, \genesis skills tap into an OpenShift-based system to manage system configuration, deployment, and life cycle of agentic validation.

\textbf{Knowledge Plane: \synapse.} The right edge of Fig.~\ref{fig:genesis-intro} shows \synapse, the persistent knowledge layer that the four horizontal layers all read from and write to. \synapse plays three roles: (i) \emph{source of ground truth}, presenting curated 3GPP and O-RAN specifications, a curated corpus of research papers, reference implementations, and lab inventory, exposed through hybrid retrieval, verified by human experts, and organized according to a telecom-specific ontology that composes a technical and institutional knowledge graph; (ii) \emph{recipient of generated knowledge}, with every \genesis run writing back code diffs with spec-to-code traceability, traces, logs, and full experimental campaigns via the \ingest stage, also behind a human-reviewer gate; and (iii) \emph{cross-capability substrate}. Here, an artifact produced by one capability (a \harden patch, an \optimize dataset, an \discover dApp) becomes an input the next capability can consume. 

\textbf{\gls{llm} Backend.} 
\genesis runs against commercial cloud-hosted models (Claude Opus~4.7 and Sonnet~4.6 in our current deployment), and against open-weight models served locally (gpt-oss, Llama~4, Phi, Gemma), with the orchestrator matching appropriate models for the tasks at hand. 

\textbf{The closed loop.} Reading the figure from top to bottom, an intent enters at the top, is routed by the orchestrator, planned and executed by specialists composing skills under hook supervision, validated on the substrate's three tiers, and the resulting traces, logs, and experiment results flow back up into \synapse and into the agents' next decision. The entire system is a closed loop in which the testbed and the knowledge plane co-evolve with the agentic framework.

\subsection{Contributions and Paper Structure}

This paper discusses the design of \genesis, rooted in the closed-loop agentic harness discussed above, and of three use cases, including the evaluation of the performance of \genesis in generating a valid output from the use-case intent, and the validation of the synthesized solution.
 
\noindent\textbf{\genesis Use Cases.} We develop three illustrative use cases for \genesis. In the first, \genesis implements the \texttt{RRC.ConnMean} \gls{kpm} from \gls{3gpp}~\gls{ts}~28.552~\cite{ts28552} in an open-source 5G stack using the \synthesize pipeline, to demonstrate the spec~$\to$~code~$\to$~\ota propagation on a single feature. The second, \gls{cho} with a closed-loop E2SM-RC xApp, is a \emph{cross-capability} example that exercises \synthesize, \testcap, and \harden together. \gls{cho} is a multi-specification feature, with the additional closed-loop control. In this case, synthesis, conformance testing, and field hardening are more effective when executed jointly. The third showcases the \discover capability, and how it connects to \synthesize and \testcap. We leverage the autonomous research loop we discussed in~\cite{allstar}, which takes high-level intents and generates new schedulers, both from an algorithmic point of view (i.e., combining different rewards and allocation policies into a comprehensive scheduler) and implementation (the scheduler is functional and deployed on the \genesis infrastructure).

\noindent\textbf{Our Contribution.}
This paper makes three key contributions, introducing \genesis as
the first agentic framework that drives the full \gls{ran}
R\&D life-cycle from intent to \ota evidence on real radios:
\begin{enumerate}

\item \textbf{The first end-to-end demonstration that an agentic framework can synthesize, test, and discover RAN functionalities on production-grade radios.} Across multiple statistically independent runs, \genesis achieves 100\% success on the implementation of the \texttt{RRC.ConnMean} \gls{kpm} and of \gls{cho} with a closed-loop E2SM-RC xApp, and maps the ALLSTaR autonomous scheduling loop~\cite{allstar} as the \discover anchor. The off-the-shelf baseline (Claude Code with Opus~4.7) produces no working implementation on any attempt for either of the first two case studies (Sec.~\ref{sec:results}).

\item \textbf{An agentic architecture that closes the RAN R\&D loop end-to-end, anchored in a staged validation continuum.} \genesis composes three portable primitives (\emph{agents}, \emph{skills}, \emph{hooks}) with a persistent knowledge plane (\synapse) into six capability pipelines that together cover the full life-cycle. The same primitives drive a three-tier validation continuum (\rfsim~$\to$~emulation~$\to$~\xfiveg) where every test outcome is routed back into the agent's next decision via hooks, which also enforce non-bypassable safety gates on actions that touch live hardware. Every artifact produced by one capability flows through the knowledge plane to become an input that compounds over time, and the primitives are portable across the major agentic runtimes (Secs.~\ref{sec:design}--\ref{sec:testbed}).

\item \textbf{A non-obvious model-selection tradeoff for agentic RAN engineering.} Per-stage profiling over statistically independent trials reveals that two of the six \synthesize stages (implementing the feature and executing the tests) dominate both cost and wall-clock, and that once cost is normalized by success rate a mid-tier \gls{llm} matches a frontier one on cost-per-successful-feature while trading wall-clock for throughput, which turns the model selection into a deployable latency/throughput decision rather than a strict cost or quality dominance (Sec.~\ref{sec:results}).
\end{enumerate}

The remainder of the paper is organized as follows.
Sec.~\ref{sec:related} surveys related work, and
Sec.~\ref{sec:motivation} maps the \genesis primitives onto the capability pipelines.
Sec.~\ref{sec:design} describes the agent/skill/hook architecture.
Sec.~\ref{sec:testbed} describes the agentic testbed and validation continuum.
Sec.~\ref{sec:design:synthesize} dives into the \synthesize pipeline.
Secs.~\ref{sec:res:case:kpm}--\ref{sec:res:case:mac} present three end-to-end case studies, including the validation of the \genesis-generated solutions.
Sec.~\ref{sec:results} reports the numerical evaluation of \genesis performance, and
Sec.~\ref{sec:conclusion} concludes the paper.

\begin{table*}[!t]
\centering
\caption{Comparison of \genesis with the closest analog in each research thread. \cmark: feature is central to the cited work; \pmark: partially supported or out of scope of the work's primary contribution; \xmark: feature is absent.}
\label{tab:related-comparison}
\renewcommand{\arraystretch}{1.2}
\setlength{\tabcolsep}{5pt}
\footnotesize
\begin{tabular}{@{}l l ccc cc@{}}
\toprule
\multirow{2}{*}{\textbf{System}} &
\multirow{2}{*}{\textbf{Primary Target}} &
\multicolumn{3}{c}{\textbf{Reasoning \& Knowledge}} &
\multicolumn{2}{c}{\textbf{Validation Reach}} \\
\cmidrule(lr){3-5} \cmidrule(l){6-7}
& & \textbf{Multi-} & \textbf{Spec-} & \textbf{Closed} & \textbf{Real} & \textbf{Spec-to-} \\
& & \textbf{agent} & \textbf{grounded} & \textbf{loop} & \textbf{radio} & \textbf{\ota} \\
\midrule
MetaGPT~\cite{hong2023metagpt}                  & Software engineering    & \cmark & \xmark & \pmark & \xmark & \xmark \\
Voyager~\cite{wang2023voyager}                  & Open-ended skill learning & \xmark & \xmark & \cmark & \xmark & \xmark \\
Glia~\cite{hamadanian2025glia}                  & GPU-cluster systems design & \cmark & \xmark & \cmark & \pmark & \xmark \\
Navidan et al.~\cite{navidan2026toward}         & O-RAN operations (runtime) & \cmark & \xmark & \pmark & \cmark & \xmark \\
% OpenRAN~Gym~\cite{bonati2020openrangym23}       & O-RAN AI/ML toolbox     & \xmark & \xmark & \xmark & \cmark & \xmark \\
Jiang et al.~\cite{jiang2026agentic}            & Intent-based networking (runtime) & \cmark & \pmark & \cmark & \xmark & \xmark \\
Ferrag et al.~\cite{ferrag2026sixg}             & Agentic AI-native 6G (architecture) & \cmark & \pmark & \pmark & \xmark & \xmark \\
Dev et al.~\cite{dev2025advanced}               & Agentic 6G architectures (V2X exp.) & \cmark & \pmark & \cmark & \xmark & \xmark \\
Gajjar \& Shah~\cite{gajjar2026agents}          & AI-RAN cognitive OS (vision) & \cmark & \pmark & \xmark & \xmark & \xmark \\
ComAgent~\cite{li2026comagent} & Wireless optimization design & \cmark & \pmark & \cmark & \xmark & \xmark \\
AI Telco Engineer~\cite{aitaoudia2026aitelco}  & PHY-algorithm design (sim.) & \cmark & \xmark & \cmark & \xmark & \xmark \\
Zota et al.~\cite{electronics14091775}              & Enterprise IT operations & \pmark & \xmark & \xmark & \xmark & \xmark \\
5GReasoner~\cite{lte-nas-fuzzing19}             & Protocol security analysis & \xmark & \cmark & \xmark & \xmark & \xmark \\
ALLSTaR~\cite{allstar} & MAC Scheduling & \xmark & \xmark & \cmark & \cmark & \pmark \\ 
AgentRAN~\cite{elkael2026agentran} & r/x/dApp-based RAN optimization & \cmark & \xmark & \cmark & \cmark & \xmark \\ 
\midrule
\rowcolor{genesisrow}
\textbf{\genesis (this work)}                   & \textbf{End-to-end RAN engineering (eng.\ life-cycle)} & \cmark & \cmark & \cmark & \cmark & \cmark \\
\bottomrule
\end{tabular}

\end{table*}

\section{Related Work}
\label{sec:related}

\genesis lies at the intersection of three research areas, discussed in the following paragraphs: multi-agent \gls{llm} frameworks, \gls{llm}-assisted software engineering, and specification-driven code generation and protocol testing. Table~\ref{tab:related-comparison} summarizes the differences between \genesis and state of the art approaches in this area.

\textbf{Multi-Agent \gls{llm} Frameworks.}
A growing body of work decomposes complex tasks across specialized \gls{llm} agents. General-purpose runtimes span conversational message passing (AutoGen~\cite{wu2024autogen}), role-specific software pipelines (MetaGPT~\cite{hong2023metagpt}, ChatDev~\cite{qian2024chatdev}), role/goal task backlogs (CrewAI~\cite{crewai2026}), state-graph orchestration (LangGraph~\cite{langgraph2026}), open-ended skill learning (Voyager~\cite{wang2023voyager}), and \glspl{sdk} such as Claude Agents~\cite{claude-agents-doc} and OpenAI Agents~\cite{openai-agents-sdk}. Zhou~et~al.~\cite{zhou2026externalization} unify these under \emph{externalization}---relocating state, procedural know-how, and interaction structure from inside the model into external memory, skills, and a coordinating \emph{harness}---which maps directly onto \genesis's agent/skill/hook split (Sec.~\ref{sec:design}). The closest non-wireless analog to \genesis, \emph{Glia}~\cite{hamadanian2025glia}, couples a reasoning/experiment/analysis loop to a simulator--emulator--testbed continuum for \gls{gpu}-cluster scheduling, paralleling our staged validation (Sec.~\ref{sec:testbed:staged}) and hooks (Sec.~\ref{sec:design:hooks}), but without integration of technical specifications or wireless components. 

A second line applies these techniques to network optimization~\cite{jiang2025tutorial}. At runtime, Navidan~et~al.~\cite{navidan2026toward} bind agent capacity to control-loop latency in \gls{oran} (with different models sizes across rApps, xApps, and \gls{ran}), while Jiang~et~al.~\cite{jiang2026agentic} build an agentic \gls{ibn} orchestrator over \gls{ran}/core specialists and observe that prompt variations induce compounding biases in the operations. This directly motivates our externalization of deterministic, procedural guidance as \emph{skills} and of policy as \emph{hooks}. At the architecture layer, Ferrag~et~al.~\cite{ferrag2026sixg} cast \glspl{llm} as bounded, policy-governed reasoning entities above deterministic \gls{3gpp} infrastructure, and Gajjar and Shah~\cite{gajjar2026agents} promote them to a ``cognitive OS''. Dev~et~al.~\cite{dev2025advanced} use three LLaMA-2 agents in crewAI with a \gls{rag} corpus (arXiv, telecom Q\&A, \gls{3gpp}) to jointly tune V2X power/modulation/retransmission in ns-3 and SUMO. Our recent AgentRAN~\cite{elkael2026agentran} work proposes a \gls{oran}-based hierarchical framework which distributes agents over rApps and xApps which control dApps in the \gls{ran}.

For numerical and PHY design, Li~et~al.'s \emph{ComAgent}~\cite{li2026comagent} coordinates Literature/Planning/Coding/Scoring agents that match expert non-AI baselines for wireless power transfer case. NVIDIA's \emph{AI Telco Engineer}~\cite{aitaoudia2026aitelco} evolves containerized agents over Sionna~\cite{hoydis2023sionna} via a leaderboard loop to design channel estimators, link adaptation, and \gls{ldpc} decoders. Both papers share with \genesis the conviction that agentic correctness must be established through execution, but differing in substrate (numerical/link-level simulation vs.\ \ota behavior on production radios).

Where prior agentic work \emph{operates} \gls{oran}~\cite{navidan2026toward,jiang2026agentic}, \emph{architects} 6G~\cite{ferrag2026sixg}, \emph{orchestrates} narrow ML~\cite{gajjar2026agents}, \emph{tunes} parameters~\cite{dev2025advanced}, \emph{designs} algorithms in simulation~\cite{li2026comagent,aitaoudia2026aitelco}, or \emph{designs} cluster schedulers~\cite{hamadanian2025glia}, \genesis \emph{engineers and evolves} the software stack itself, propagating features from specification text through code, compilation, deployment, and \ota validation, with abstractions that can propagate to multiple runtime environments.

\textbf{\gls{llm}-Assisted Software Engineering.} A second thread treats \glspl{llm} as software-engineering assistants: SWE-bench~\cite{jimenez2023swe} and derivatives benchmark repository-scale bug fixing, SWE-agent~\cite{yang2024swe} designs the agent-computer tool surface, and self-repair/planning techniques~\cite{self-repair24} iterate on failing tests. \genesis shares the iterate-to-acceptance philosophy, but its acceptance criterion is not a test suite---it is validation on real radio infrastructure, a gap that motivates our staged-validation continuum (Sec.~\ref{sec:testbed:staged}) and the policy/audit role of hooks (Sec.~\ref{sec:design:hooks}). AI5GTest~\cite{ganiyu2025ai5gtest} leverages three \glspl{llm} to generate tests, validate stack functionality, and debug failures. Compared to this, \genesis focuses on a more generic agentic approach, where different \glspl{llm} are autonomously selected by agents according to their functionality and needs, and on the sythesis of protocol stack features, besides testing harness.

\textbf{Specification-Driven Code Generation and Protocol
Testing.}
% }
% \label{sec:rel:specgen}
%
A third thread bridges specifications and code directly, either for analysis or implementation. Early work on machine-readable \gls{3gpp} specifications~\cite{achine-readable-3gpp24} explored structured clause representations, while structured fuzzing of LTE/\gls{5g} \gls{nas}~\cite{lte-nas-fuzzing19} discovers conformance gaps through test generation. More recently, \glspl{llm} have been shown as effective in translating \glspl{rfc} into code~\cite{llm-rfc-codegen23}. \genesis generally deals with more complex 3GPP and O-RAN specifications, which are usually an order of magnitude longer compared to IETF \glspl{rfc}. 
Multiple foundational models and test benches related to RAN specifications have also been proposed~\cite{gajjar2025ORANSight,gajjar2026teleresiliencebench,rezazadeh2025experimental}. Rather than focusing on a foundational model, \genesis develops agentic components that can leverage different \glspl{llm} and test benches.

\textbf{Summary.}
\label{sec:rel:position}
To our knowledge, \genesis is the first end-to-end framework that (i)~applies multi-agent \glspl{llm} to the full life-cycle of RAN software generation and testing, including real-world over-the-air deployment, (ii)~exposes the design as portable agent/skill/hook primitives with an explicit cross-runtime mapping, and (iii)~addresses the end-to-end R\&D life cycle, across multiple open-source repositories, from specification text to \ota validation on real radios. Table~\ref{tab:related-comparison} compares \genesis with the closest analog in each thread along five axes (multi-agent reasoning, spec-grounded outputs, closed-loop validation, real-radio reach, and spec-to-\ota propagation); \genesis is the only system combining all five.

\section{\genesis Capabilities}
\label{sec:motivation}

\begin{figure}[!t]
  \centering
  \includegraphics[width=\linewidth]{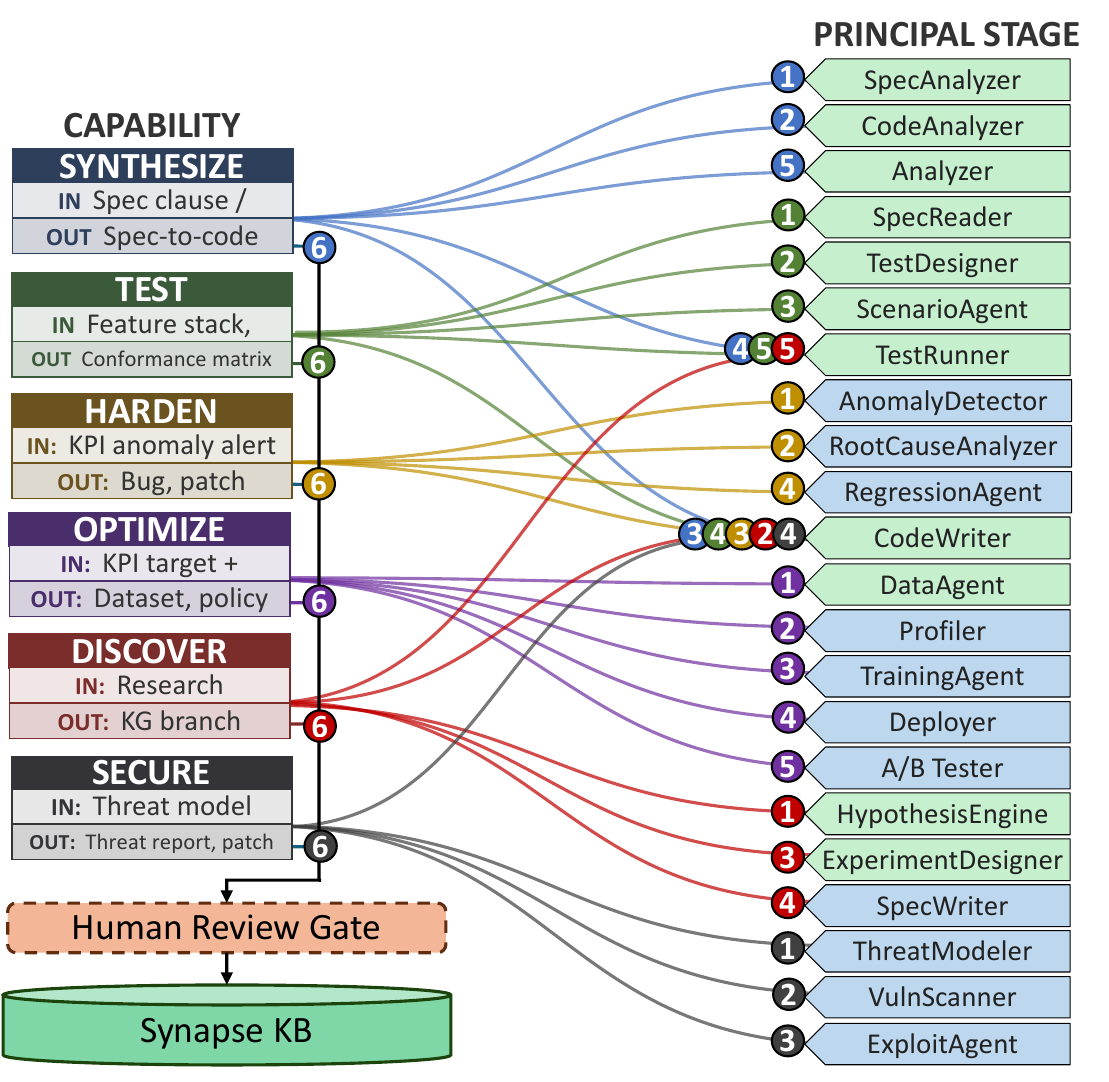}
  \caption{Summary of the six \genesis capabilities, including the input and output for each one, the stages in which they unfold (in the order indicated by the numbers, color-coded to the capability), and the ingestion of the outcome in the knowledge base, after a human review.
  }
  \label{fig:design:genesis_capabilities}
\end{figure}

As shown in Fig.~\ref{fig:genesis-intro}, \genesis provides a coordination and reasoning
layer that turns a federation of specifications, testbeds, and operator
infrastructure into a coherent, \emph{autonomous} agentic engineering engine, organized in six autonomous capabilities (i.e., \synthesize, \testcap, \harden, \optimize, \discover, and
\secureit). Figure~\ref{fig:design:genesis_capabilities} connects the capabilities (and their input/output relationship) to the \genesis agents, and illustrates how they share a common step: after a human review gate, the outcome of each pipeline is ingested in the knowledge base. Next, we summarize the role of each capability.

\paragraph*{\synthesize}\label{sec:usecases:synthesize}
A spec-to-code pipeline. Given a \gls{3gpp} or O-RAN clause, or a research
paper, an orchestrator drives a six-stage pipeline that queries the knowledge base, maps requirements onto the stack codebase,
writes the change, validates it across the tiered continuum, and
ingests the result.

\paragraph*{\testcap}\label{sec:usecases:test}
Regression-grade conformance testing against system and standard specifications. The pipeline generates test cases,
configures testing scenarios, runs
the full campaign on simulation/Colosseum/\xfiveg, and flags regressions
against spec-derived acceptance criteria. The knowledge-base artifact
is a conformance pass/fail matrix indexed by specification section
plus a reproducible scenario bundle.

\paragraph*{\harden}\label{sec:usecases:harden}
Bug-to-fix loop. An anomaly detector watches \gls{kpm} counters
(e.g., a \gls{harq} retransmission spike under high \gls{ue} load), a
root-cause analyzer traces the anomaly to a code path, a
\codewriter produces a targeted patch, a \testrunner validates under
the same conditions that triggered the bug, and a \regressionAgent
runs the full regression suite before the patch reaches production.

\paragraph*{\optimize}\label{sec:usecases:optimize}
Data-driven adaptation and sim-to-real transfer. An instantiation of this workflow is the AI-RAN data factory discussed in our prior work~\cite{elkael2026agentran}.
An example
is learning \gls{cqi}-to-\gls{mcs} mapping: Colosseum emulation campaigns
generate a dataset, a training agent fits an \gls{ml} policy, a
deployer pushes it as an xApp/rApp via the non-RT~\gls{ric},
and an A/B tester evaluates against the baseline on \xfiveg with
statistical significance.

\paragraph*{\discover}\label{sec:usecases:evolve}
From research hypothesis to implementation and experimental validation, with paper or standard contribution. A researcher elaborates an 
hypothesis to be developed and tested (e.g., a learned waveform-adaptation
scheme for \gls{6g}~\cite{oshea2017dlphy}); a \codewriter implements a candidate feature on
a \gls{6g} branch; a \testrunner produces emulation and \ota
evidence; a \specwriter drafts the corresponding \gls{3gpp} document or
change request bound to that evidence.

\paragraph*{\secureit}\label{sec:usecases:secure}
Adversarial security analysis of \gls{ran} signaling paths
(\gls{rrc}, \gls{nas}, M-Plane) with non-bypassable safety gates. A
threat modeler applies the STRIDE taxonomy~\cite{hernan2006stride} to the relevant security specifications (\gls{ts}~33.501/511); a vulnerability scanner
combines static analysis and protocol fuzzing; attacks run only in a
sandboxed Colosseum instance; a \codewriter generates patches that must pass conformance regression. Policy enforcement, rollback triggers,
and audit trails are implemented as hooks
(Sec.~\ref{sec:design:hooks}).

\smallskip
Although their goals differ, the six capabilities share the same
architectural skeleton---orchestrator, specialists, staged validation,
Gatekeeper---because they are the structural consequence of the
design primitives presented next. 

\section{System Architecture}
\label{sec:design}

\genesis rests on a small number of primitives chosen to be \emph{composable}, \emph{observable}, and \emph{portable} across different agentic frameworks. In this section, we first describe the agent/skill/hook triad and their orchestration (Secs.~\ref{sec:design:agents}--\ref{sec:design:orchestration}). We then formalize inventory as code~(Sec.~\ref{sec:design:inventory}), the knowledge base (Sec.~\ref{sec:design:kb}), and discuss how the abstraction maps onto the major agentic runtime available today~(Sec.~\ref{sec:design:portability}). Figure~\ref{fig:genesis-intro} summarizes the architecture.

\subsection{{Agents: Reasoners with a Tool Surface}}
\label{sec:design:agents}

Classically, an \emph{agent} is an entity that perceives its environment through sensors, maintains internal state, and acts on the environment through actuators in service of a goal\cite{russell2020aima}. The present generation of \gls{llm}-based agents~\cite{claude-agents-doc} follow the same pattern: an LLM serves as the reasoning core, its context window as internal state, and tool calls as both sensors and actuators. We define an agent in \genesis as a triple: \emph{(persona, tool surface, control loop)}. The \emph{persona} describes the nature of the agent: its identity, expertise, decision authority, constraints, and preferred reasoning style. This makes it possible to create the \emph{specialists} introduced in Fig.~\ref{fig:genesis-intro} and Sec.~\ref{sec:introduction}, i.e., agents with different personas, to drive the autonomous loop, while separating concerns and keeping a manageable context. The \emph{tool surface} is the set of actions the model may take, together with their input schemas. For example, a DevOps specialist is equipped with information on how to interact with system automation. The \emph{control loop} is a ReAct-style alternation of thought and action~\cite{yao2022react}: at each step the model emits a natural-language reasoning trace (the \emph{thought}) followed by a single tool invocation (the \emph{action}). The runtime executes the call and appends its result (the \emph{observation}) to the context, and the model is re-invoked to produce the next thought. The loop terminates when the model emits a final response in place of an action, or when a runtime budget (tokens, wall time, step count) is exhausted.
 
Concretely, agents in \genesis are represented as markdown files and are referred to as specialists. Each file contains at minimum a \texttt{description}, and a list of \texttt{tools/skills} (Sec.~\ref{sec:design:skills}) the specialist may invoke. The \texttt{description} field is the routing grammar with which the orchestrator decides when to dispatch a specialist. A specific, action-oriented \texttt{description} (``returns an \texttt{EXPERIMENT\_PLAN} for the \xfiveg testbed given constraints'') yields reliable routing while a vague one (``knows about \xfiveg'') does not. Examples of specialists included in \genesis are:
\begin{itemize}
    \item a DevOps specialist, which manages OpenShift
and pod lifecycle;

    \item the RAN specialist, taking care of building, configuring, and running the \gls{ran} stack (\gls{oai}, in the first \genesis implementation);

    \item the Radio specialist, which can configure O-RAN \glspl{ru} (e.g., a Foxconn \gls{ru} part of \xfiveg) via SSH/M-Plane;

    \item a UE specialist, which can interact with Sierra~Wireless \glspl{ue} and \gls{oai} softUE deployed in \xfiveg;

    \item the Testbed specialist, which covers planning of experiments and tests on the \ota or emulated testbeds.
\end{itemize}
Additional specialists can include stage-specific agents, e.g., an xApp specialist to interact with the O-RAN Near-RT RIC, and KPM specialists to analyze performance evaluation. 

\subsection{{Skills: Agent-Invoked Instruction Packages}}
\label{sec:design:skills}

Agents reason using \glspl{llm}. To \emph{execute} deterministic, procedural steps, the agent's persona file directs it to invoke a \emph{skill} whose own body in turn may delegate to a script. For complex, frequent, and scriptable procedures, the agent chooses \emph{what} to do and \emph{when}, the skill specifies \emph{how}. This grounds the agent in precise capabilities, avoiding the need to re-think through common procedures, and increasing the likelihood that an intent is successfully implemented.

A \genesis skill is also represented as a markdown file, i.e.,  \texttt{SKILL.md}. The body of the file is adaptive ``guidance'', meaning that the specialist may skip steps, reorder them, or improvise. For any step where that behavior would be unacceptable, or that is completely deterministic, the skill's markdown  delegates to a script and instructs the specialist to invoke it with a specific argument shape. This layering keeps the interpretive surface (agent reasoning) separate from the deterministic one (script execution). Thanks to this design choice, each surface can be reviewed and hardened independently, and specialist regressions on one surface do not silently rewrite the other. It also avoids compounding of probabilistic errors generated by the stochastic nature of the agents reasoning.

\begin{figure}[!t]
  \centering
  \includegraphics[width=0.78\columnwidth]{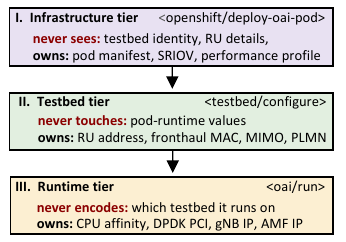}
  \caption{Tiered skill invocation chain. Each tier owns a disjoint set of inputs. Tiers above never see what tiers below depend on.}
  \label{fig:design:three_tier_skills}
\end{figure}

Fig.~\ref{fig:design:three_tier_skills} illustrates a concrete example of composition for running a \gls{gnb} on \xfiveg. It is realized as a chain of three skills, each in its own \texttt{SKILL.md} and specific elements that are within scope of the skill. Each one is invoked only when its inputs change, i.e., when the agent decides to update the parameters that are within the scope of the skill. This minimizes the range of commands that need to be run for each agentic action and increases the speed of convergence for \genesis flows.

\subsection{{Hooks: Observability, Policy, and Audit}}
\label{sec:design:hooks}

Agents and skills specify \emph{what} is done, whereas hooks specify what
happens \emph{around} every action. Hooks execute deterministically and out-of-band from the reasoning loop, do not consume context tokens, and are invariant to the specialist or skill in scope at the moment they fire. The portability of hooks is the weakest of the three primitives, since the set of available events and the contract for blocking an action vary by runtime. A Claude Code implementation, for instance, currently exposes 29 such events, of which \genesis subscribes to five: \texttt{UserPromptSubmit},
\texttt{PreToolUse}, \texttt{PostToolUse}, \texttt{Notification},
and \texttt{Stop}~\cite{claude-hooks-doc}. These five suffice to
anchor three cross-cutting concerns.

\begin{itemize}
\item \textbf{Observability.} The agent's context window is a finite resource whose consumption competes directly with reasoning capacity, which makes in-loop telemetry collection incompatible with the duration of an end-to-end \synthesize run that may exceed an hour and span dozens of pod-level interactions. Hooks resolve this contention by externalizing observability. For example, a \texttt{PostToolUse} hook serializes each tool call and its result to a structured event log.

\item \textbf{Policy.} Hooks are also a \emph{safety plane} and a way to enforce policies. When autonomous agents interact with physical RF infrastructure, deterministic safety guardrails are mandatory. Hooks act as a non-bypassable policy plane. By binding a hook to a given event, \genesis can intercept and evaluate an action's payload \emph{before} it touches the testbed. If a specialist attempts an unsafe operation the hook terminates the call and returns a hard block (e.g., \texttt{ {"decision":"block"}}).

\item \textbf{Audit.} Features destined for production networks require strict provenance. Because hooks run as ordinary shell commands outside the LLM's control, they serve as an objective witness to the pipeline. They capture the exact sequence of tool calls, code diffs, and testbed configurations. This ensures that every action \genesis takes could be audited.

\end{itemize}

\begin{table}[!t]
\centering
\caption{The \genesis primitive triad.}
\label{tab:triad}
\footnotesize
\begin{tabular}{llll}
\toprule & \textbf{Agent} & \textbf{Skill} & \textbf{Hook} \\
\midrule Plane & Reasoning & Execution & Observability \\
Behavior & Adaptive & Guided & Reactive \\
Artifact  & \texttt{.md} persona & \texttt{.md} $+$ script & Event binding \\
Example & Specialists & \texttt{oai/run} & Gatekeeper gate \\
\bottomrule
\end{tabular}
\end{table}

\subsection{Orchestration}
\label{sec:design:orchestration}

Overall, the combination of agents, skills, and hooks constitutes the \genesis primitive triad summarized in Table~\ref{tab:triad}. Each component covers different functionality planes, comes with different behaviors, and is represented by different artifacts. On top of this triad, an orchestrator (shown in Fig.~\ref{fig:genesis-intro}) is implemented by the specific agentic framework adopted in \genesis (e.g., Claude Code in this first iteration, see Sec.~\ref{sec:design:portability}). The orchestrator coordinates the activities of the triad. The hierarchy is as follows: an orchestrator may dispatch specialists, and specialists in turn rely on skills (but not on other specialists). In our implementation, the parent conversation acts as the orchestrator and specialists cannot dispatch one another by design. Therefore, only the orchestrator dispatches the specialists, preserving a clear call graph (also shown in Fig.~\ref{fig:design:genesis_capabilities}). Specialists are run on models selected by the orchestrator for reasoning depth: %Opus-class
top-tier (e.g., Opus) for orchestration and for complex reasoning, mid-class models (e.g., Sonnet) for specialists. Cost and latency are thus managed explicitly, as we discuss in Sec.~\ref{sec:results}. 

Further, \genesis agentic pipelines halt on failure and surface logs rather than attempting autonomous recovery across stages, allowing a human review of evidence before resumption. This is both a guardrail (autonomous cross-stage recovery would risk compounding errors) and a design choice with human review as quality gate. Within a stage, bounded local retry loops are permitted.

\begin{figure}[b]
\begin{lstlisting}[caption={Distance matrix of the \xfiveg testbed inventory. The file is the single source of truth read by the specialist to resolve parameters at planning time.},label={lst:inventory}]
- sierra_ue
    - plmn: "00105"
    - foxconn01:
          distance: close
          avg_rsrp_dBm: -75
          ru_attn_dB: 10
    - foxconn02:
          distance: far
          avg_rsrp_dBm: -110
          ru_attn_dB: 10
- samsung_ue
\end{lstlisting} 
\end{figure}

\begin{table*}[!t]
\centering
\caption{Mapping the \genesis primitives onto contemporary agentic runtimes.}
\label{tab:portability}
\resizebox{\textwidth}{!}{%
\begin{tabular}{@{}lp{3.8cm}p{2.8cm}p{3.8cm}p{5.0cm}@{}}
\toprule
\textbf{Runtime}
  & \textbf{Agent (persona)}
  & \textbf{Skill (procedure)}
  & \textbf{Hook (event)}
  & \textbf{Orchestrator} \\
\midrule
AutoGen~\cite{wu2024autogen}
  & \texttt{AssistantAgent}
  & \texttt{FunctionTool}
  & \texttt{@message\_handler}
  & \texttt{Swarm} \\
CrewAI~\cite{crewai2026}
  & Role-based \texttt{Agent}
  & \texttt{BaseTool}
  & \texttt{BaseEventListener}
  & \texttt{Crew} with \texttt{Process.sequential} \\
LangGraph~\cite{langgraph2026}
  & \texttt{StateGraph} node 
  & \texttt{ToolNode}
  & \texttt{BaseCallbackHandler}
  & Supervisor \texttt{StateGraph} \\
OpenAI Agents SDK~\cite{openai-agents-sdk}
  & \texttt{Agent} class
  & \texttt{@function\_tool}
  & \texttt{RunHooks} / \texttt{AgentHooks}
  & \texttt{Agent} with \texttt{handoffs}; \texttt{Runner.run()} \\
Claude Code~\cite{claude-agents-doc,claude-hooks-doc}
  & \texttt{.claude/agents/*.md} 
  & \texttt{SKILL.md}
  & \texttt{.claude/settings.json}
  & Parent session \\
\bottomrule
\end{tabular}}
\end{table*}

\subsection{{Inventory as Code}}
\label{sec:design:inventory}
 
All hardware, network, and device information lives in YAML files and is consumed by relevant specialists at planning time. For each testbed, we define a list of available \glspl{ru}, \glspl{ue}, core networks and a distance matrix that records measured \gls{rsrp} values between \gls{ue} and every reachable \gls{ru}~(Listing~\ref{lst:inventory}). This information is then ingested by relevant specialists on demand.

This design keeps the specialists themselves stable as the lab or deployment environments evolve: moving a \gls{ue} between rooms, provisioning a new network slice, or replacing a failed \gls{ru} changes a YAML entry rather than any specialist logic. It also enables reproducibility: the exact physical state of the lab at the time of an experiment can be serialized into the knowledge base together with the experimental results.

\subsection{The Knowledge Base: \synapse}
\label{sec:design:kb}

Beyond autonomous reasoning/action loops, \genesis relies on a long-term memory for ground truth (technical specifications, papers, documents) and \genesis' output. As shown in Fig.~\ref{fig:design:genesis_capabilities}, agents within the \genesis capabilities query the knowledge base to retrieve information, and the outcome of \genesis' experiments is stored in \synapse. Therefore, this represents the long-term memory of the framework, and what provides connectivity and shared knowledge across the \genesis capabilities. \synapse is based on an ingestion pipeline that maps information into a vector and keyword index (hybrid \texttt{pgvector}/BM25), and builds a knowledge graph based on a rich, human-generated ontology. This ontology describes complex relationships across the \gls{ran} technical domain, as well as institutional notions related to standardization, commercialization, and the general telecom ecosystem. In the context of \genesis, \synapse plays three roles simultaneously:

\begin{itemize}
\item \textbf{Source of ground truth.} \synapse hosts the corpus that every \specanalyzer and \codeanalyzer step queries: ingested 3GPP and O-RAN technical specifications, reference open-source implementations (\gls{oai}\cite{kaltenberger2019openairinterface}, srsRAN\cite{srsran16}\cite{ocudu}, OSC RIC\cite{osc-ric}, FlexRIC~\cite{flexric}), prior experimental campaigns, and the lab inventory described in Sec.~\ref{sec:design:inventory}. Crucially, this corpus is expert-verified rather than randomly scraped from the open web: human curators sign off on each ingested specification version, so that downstream agents can ground their outputs in artifacts whose provenance is auditable. The knowledge graph is based on a curated ontology that describes technical and institutional relationships in the telecom ecosystem.

\item \textbf{Recipient of generated knowledge.} Every \genesis run terminates with the \ingest stage writing back to \synapse (Fig.~\ref{fig:design:genesis_capabilities}): code diffs with spec-to-code traceability (measurement name~$\to$~specification clause~$\to$~source files and line numbers), gNB and xApp logs, E2 indication traces, \analyzer verdicts, and complete experimental campaigns including the inventory snapshot used.

\item \textbf{Cross-capability substrate.} The artifacts that one capability produces become the input that another consumes. A \harden bug-fix patch enriches the regression suite that \testcap re-runs on every future change; an \optimize training dataset and the trained policy that ships with it become a reusable benchmark for the next \optimize run; a \discover result (novel dApp~\cite{lacava2025dapps}, evaluation campaign, paper draft) becomes a candidate feature for a future \synthesize run when the corresponding clause is standardized.
\end{itemize}

\subsection{{Portability Across Agentic Frameworks}}
\label{sec:design:portability}

A legitimate concern with any agentic system is lock-in to a specific runtime and/or \gls{llm}. This would prevent adopting state-of-the-art reasoning tools whose performance changes in a matter of weeks. The \genesis primitives are designed to be portable. Our implementation targets the most capable agentic systems available at the time of writing, but the underlying directives (specialist personas, \texttt{SKILL.md} files, and orchestration prompts) are expressed entirely in markdown and natural language. Substituting the underlying agent therefore reduces to mapping agent-specific entry points onto the same directives, rather than rewriting the workflows. Table~\ref{tab:portability} maps the abstraction onto the major frameworks in use today.

Three properties make the abstraction portable in practice. First, markdown "specialist" personas translate to the system prompt and tool list that every agentic runtime consumes, the only adapter required is a loader that parses the frontmatter into the framework's agent-selection mechanism. Second, \texttt{SKILL.md} files delegate to scripts rather than embed logic, so porting a skill between frameworks is a matter of wrapping the same script with a framework-specific tool declaration. Third, modern agentic frameworks are converging on shared conventions for top-level instruction files (e.g., \texttt{AGENTS.md}), which further reduces the cost of moving a \genesis deployment across runtimes to a mapping between conventions rather than a re-implementation of workflows.

\section{The Agentic Testbed}
\label{sec:testbed}

\begin{figure}[!t]
 \centering  \includegraphics[width=\linewidth]{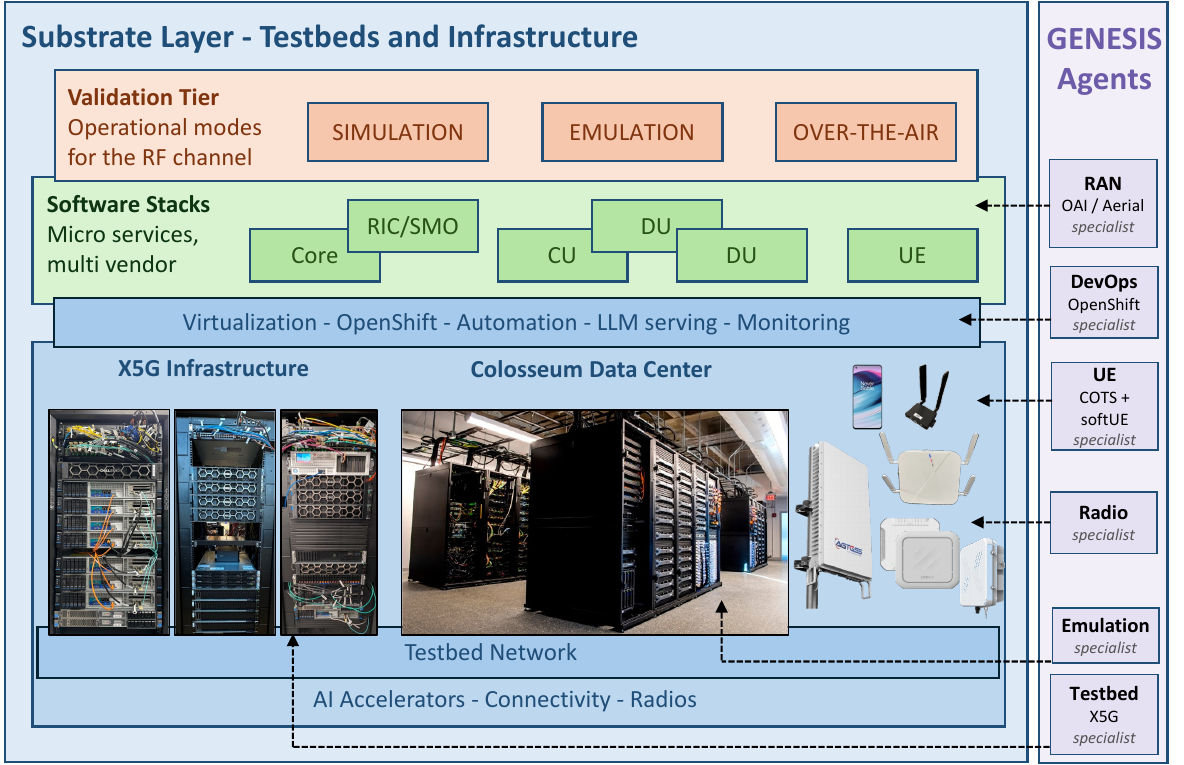}
  \caption{\genesis agentic testbed and relationships between specialist agents and testbed components.}
  \label{fig:design:genesis_x5g}
\end{figure}

The \genesis capabilities presented in Secs.~\ref{sec:motivation}-\ref{sec:design} are only as useful as the physical and software infrastructure they can actually drive. Agents need a rich, closed-loop harness in which to test implementations, research hypotheses, and bug fixes. Figure~\ref{fig:design:genesis_x5g} shows the resulting substrate as a layered stack: compute and testbeds at the bottom (Sec.~\ref{sec:testbed:hw}), a virtualization and automation plane carrying the AI-RAN software stacks above it (Sec.~\ref{sec:testbed:sw}), system-level capabilities that connect each layer to the GENESIS specialist agents on the right of the figure (Sec.~\ref{sec:testbed:caps}), and an orthogonal three-tier validation continuum that selects the operational mode for the RF channel on every run (Sec.~\ref{sec:testbed:staged}).

\subsection{{Compute and Testbeds}}
\label{sec:testbed:hw}

The bottom band of Fig.~\ref{fig:design:genesis_x5g} aggregates two co-located physical testbeds (\xfiveg~\cite{villa2025x5g} and Colosseum~\cite{bonati2021colosseum}), together with their AI accelerators, fronthaul/connectivity, and radios, exposed to \genesis through a shared testbed network.

\xfiveg is the primary \ota target: a multi-campus private \gls{5g} spanning two buildings on separate Northeastern campuses (Boston and Burlington, MA) connected through a campus network, giving a single logical deployment that mirrors the multi-site character of commercial private \gls{5g} rollouts~\cite{villa2025x5g}. The Boston side hosts the production compute: six SMC~GH200 nodes and eight Gigabyte E251-U70 servers (\gls{gpu}-equipped) alongside Dell~R750/R760/XR5610 CPU servers (their grouping into OpenShift node classes is described in Sec.~\ref{sec:testbed:sw}). Eight Foxconn RPQN-series \glspl{ru} (n78/n48/n77) plus VVDN/LITEON/Benetel/Eridan radios cover indoor cells, and a Qulsar~QG2 \gls{ptp} grandmaster disciplined by GPS provides synchronization. The Burlington side extends this with outdoor Foxconn, Amplitech, Solid, Airspan, and Benetel \glspl{ru}. Endpoints include thirteen Sierra~Wireless \gls{cots} modems (some dual-SIM), OnePlus/iPhone/Samsung handsets, and \gls{oai} softUE instances.

Channel emulation sits between simulation and \ota, still relying on hardware in the loop, and has three components, which share infrastructure with \xfiveg but can operate independently, from a logical point of view. (i)~Colosseum~\cite{bonati2021colosseum} is a large-scale channel emulator with software-defined radio connected across a matrix of FPGA-emulated RF channels modeled on real-world propagation scenarios. (ii)~Keysight \texttt{RuSIM}/\texttt{UeSIM}/\texttt{CoreSIM} provide instrument-grade \gls{ue}, \gls{ru}, and core-network emulation for standards-conformant protocol and interface stress tests. (iii)~A custom, single-\gls{ru}, three-\gls{ue} hardware-in-the-loop emulation platform uses using the same \gls{ru}/\gls{ue} hardware family as the \xfiveg testbed, but with emulated rather than \ota channels.

\subsection{{Virtualization, Automation, and Software Stacks}}
\label{sec:testbed:sw}

Above the physical layer, Fig.~\ref{fig:design:genesis_x5g} shows a cloud-native plane (virtualization, OpenShift, automation, LLM serving, monitoring) that schedules the AI-RAN software stacks on top of it.

All workloads run as OpenShift pods scheduled on worker nodes labeled by hardware class, based on the AutoRAN framework~\cite{maxenti2025autoranautomatedzerotouchopen}, or as Colosseum LXC containers. CPU nodes with SR-IOV \glspl{nic} carry 7.2-split and CPU-based \glspl{gnb} (\gls{oai} and OCUDU stacks); \gls{gpu}-equipped nodes carry GPU-accelerated \glspl{du} (NVIDIA Aerial), with \gls{mig} slicing each accelerator across pods; and general-purpose x86 nodes carry spectrum sensing, automation, and OpenShift services. Each class of software services is mapped to an OpenShift performance profile that pins isolated \gls{cpu} cores, configures hugepages and the real-time kernel, and declares SR-IOV resource pools.

The AI-RAN stacks scheduled on this plane (the green tier in Fig.~\ref{fig:design:genesis_x5g}) integrate:

\begin{itemize}
\item \textbf{NVIDIA Aerial cuBB}~\cite{aerial} for the \gls{ai}-accelerated L1, connected via the \gls{fapi} over shared memory to the \gls{oai} L2 in a multi-container pod.
\item \textbf{\gls{oai}}~\cite{oai}, built in one of three profiles---7.2 (Open Fronthaul), monolithic (e.g., for \rfsim), or aerial (Aerial \gls{fapi})---depending on the target hardware path. \textbf{OCUDU}~\cite{ocudu} provides a monolithic or 7.2 gNB.
\item \textbf{Open5GS} for the \gls{5g} core, with twelve independent deployments (different PLMN IDs) segmented across OpenShift namespaces, enabling core-network isolation between experiments.
\item \textbf{\gls{osc} near-RT \gls{ric}}~\cite{osc-ric} and \textbf{FlexRIC}~\cite{flexric} for xApp execution, together with a \gls{smo} layer for non-RT \gls{ric}/rApp hosting and \gls{ai}/\gls{ml} pipelines.
\item \textbf{M-Plane and NETCONF/YANG clients} for Open Fronthaul \gls{ru} configuration, exposed to agents through a dedicated MCP server.
\item \textbf{Support services}: ClickHouse for in-pod time-series capture alongside cuBB, and a WebSocket daemon for scripted Sierra-\gls{ue} control.
\end{itemize}

\subsection{{System-Level Capabilities Exposed to GENESIS Specialists}}
\label{sec:testbed:caps}

The dashed lines on the right of Fig.~\ref{fig:design:genesis_x5g} connect each \genesis specialist (Sec.~\ref{sec:design:agents}) to the layer it acts on. The following paragraphs describe the system-level capability behind each of those edges, which specialists compose through skills (Sec.~\ref{sec:design:skills}) into the end-to-end experiments that back the \testrunner stage.

\paragraph*{DevOps specialist---cluster and pod lifecycle} A two-phase deployment contract ({generate}~$\to$ {apply}) produces reviewable OpenShift manifests before any workload is scheduled. Skills manage deploy/scale/delete operations, inspect node affinity, bind SR-IOV resources, claim \gls{gpu} slices, and stage performance profiles. This is the loop invoked during \synthesize's infrastructure bring-up.

\paragraph*{Radio specialist---\gls{ru} configuration and health} The \xfiveg \glspl{ru} are configured either through Open Fronthaul M-Plane (NETCONF/YANG) or through an SSH-based fallback that applies XML patches directly. Each configuration change returns a confirmed \gls{ru} MAC, fronthaul sync status, and \gls{ptp} lock state---the information \testrunner needs to validate that the radio is ready before launching the \gls{gnb}.

\paragraph*{RAN specialist---\gls{gnb} bring-up and observability} The \gls{oai} \gls{gnb} startup is driven by a set of skills that patch the configuration file according to the desired deployment conditions, rebuild and launch the process, and monitor the log for a fixed milestone sequence (\texttt{Initializing}~$\to$~\texttt{NGAP connected}~$\to$~\texttt{Cell Active}~$\to$~\texttt{PRACH received}), which brings the \gls{gnb} from inactive to connected to core and radio and a \gls{ue} connection attempt. Milestone successes or failures are presented as structured status back to the \genesis orchestrator.

\paragraph*{UE specialist---orchestration and traffic} Sierra Wireless \gls{cots} modems are controlled through a WebSocket daemon exposing connect/disconnect, SIM-slot switch, PDU-session lifecycle, ping, and iperf operations. SoftUEs are driven through the same skill surface with different APIs. An inventory-maintained \gls{ue}-\gls{ru} proximity matrix (measured \gls{rsrp}) lets the testbed and emulation specialists select physically reasonable \gls{ue}-\gls{ru} pairs for the \synthesize \ota validation runs (Sec.~\ref{sec:design:inventory}).

\paragraph*{Testbed and Emulation specialists---isolation, concurrency, and reproducibility} The system provides access to Open5GS network functions with dedicated PLMNs; subscriber provisioning and per-namespace log inspection are wrapped as skills, giving \synthesize an isolated control-plane slice per experiment. Because pods, \glspl{ru}, cores, and \glspl{ue} are parameterized by inventory and isolated by namespace and SR-IOV virtual function, multiple \xfiveg \ota and channel-emulation experiments can run concurrently with no shared-resource contention---\genesis can validate one feature on the channel-emulation tier while a previous one is still being regressed \ota. The lab state itself lives in an inventory YAML file (Sec.~\ref{sec:design:inventory}), so the exact physical configuration at the time of a \synthesize run is serialized and ingested into the knowledge base alongside the result, closing the provenance loop. The same specialists drive the channel-emulation and \ota tiers through \texttt{configure} and \texttt{experiment} skills that follow the same contract and parameter schema used in simulation.

\subsection{{Staged Validation Continuum}}
\label{sec:testbed:staged}

The top band of Fig.~\ref{fig:design:genesis_x5g} marks the three operational modes for the RF channel that \testrunner targets in sequence---simulation, emulation, \ota---each trading speed for fidelity and catching a different class of defect. \textit{Simulation} runs \rfsim (for \gls{oai}) or ZMQ with the \gls{oai} \gls{ue} (for OCUDU) on a single \gls{gnb}-\gls{ue} pair in seconds to minutes, catching compilation errors, ASN.1 encoding bugs, E2-setup misconfigurations, and logical defects. \textit{Emulation} runs the same binaries on Colosseum, Keysight, or the HIL platform in minutes to hours, catching fronthaul timing violations, \gls{fapi} interoperability bugs, real-time scheduler constraints, and multi-\gls{ue} protocol edge cases that pure software cannot reproduce. \textit{Over-the-air on \xfiveg} is the ground-truth tier, exercising real RF (antennas, fronthaul timing sensitivity, channel dynamics), scale behavior (multi-\gls{ue} iperf, mobility), and the production \gls{du}/\gls{ru} hardware path.

\section{From Primitives to Capabilities: \synthesize}
\label{sec:design:synthesize}

As suggested by Figs.~\ref{fig:genesis-intro} and~\ref{fig:design:genesis_capabilities}, the agent/skill/hook primitives, running on agentic testbed of Sec.\ref{sec:testbed}, compose into a full capability. In this section, we describe an example of such composition by focusing on \synthesize. This, together with \harden, \testcap, and \discover, is then showcased end-to-end in Secs.~\ref{sec:res:case:kpm}---\ref{sec:res:case:mac}. 

Given a published specification and a target measurement or feature name, \synthesize produces a validated change to a production-grade O-RAN stack. The same stages and skills can be leveraged by \testcap, \harden, and \optimize. 

As shown in Figure~\ref{fig:design:genesis_capabilities}, \synthesize decomposes the spec-to-\ota path into six sequential stages, each owned by a specialist agent. \textbf{(1)~\specanalyzer} grounds the work in a specification: it queries the \genesis knowledge base \synapse using the \texttt{synapse-retrieve} skill (Sec.~\ref{sec:design:kb}) and dispatches a \texttt{SpecResearcher} that uses a \texttt{spec-mapping} skill to understand the practical components associated to translating the specification into code.
For example, for \glspl{kpm}, it infers the \gls{kpm} report style and node scope~\cite{feraudo2026xdevsm}. Every field written into
\texttt{/specs/<name>.md} must be grounded in a retrieved chunk; otherwise the agent is instructed to escalate to the human operator. This is to ensure that the starting point for the implementation is actually based on real specifications.

\textbf{(2)~\codeanalyzer} maps the specification onto the target O-RAN codebase (\gls{oai} in our experiments; the pipeline also applies to other stacks, e.g.,\ OCUDU) and emits an \texttt{IMPLEMENTATION\_PLAN}. Continuing with the example of KPM features, \codeanalyzer queries two subagents, i.e., \texttt{KPM-Implementer}, which classifies the \gls{kpm} into one of four canonical O-RAN patterns (instantaneous scalar, cumulative counter with delta, ratio/percentage, histogram with label bins), and a \texttt{KPM-E2Advertiser}, which is responsible for the identification of the fields necessary to the E2 Setup. The plan is gated by human approval before the next step.

\textbf{(3)~\codewriter} implements the change. A RAN specialist agent drives the \texttt{compile} and \texttt{build} skills, iterating against the compiler log until success or a bounded retry budget is exhausted; each specialist verifies its own changes before the next begins.

\textbf{(4)~\testrunner} validates the implementation along the three-tier continuum of Sec.~\ref{sec:testbed:staged} (\rfsim, then channel emulation, then \ota on \xfiveg), reusing the same branch, configuration template, and skill set at every tier.

\textbf{(5)~\analyzer} is a generic metric extractor: given a list of metrics (name, extractor, aggregation, threshold) and a report path, it parses \testrunner's artifacts into a pass/fail verdict and a structured findings report.

\textbf{(6)~\ingest} closes the knowledge loop by staging every pipeline artifact (code diff with spec-to-code traceability map, logs, E2 trace, \analyzer verdict, Gatekeeper signature) into \synapse, after a human approval gate. Future runs retrieve these artifacts through the same \texttt{synapse-retrieve} skill used at \specanalyzer time.

Two boundaries deserve emphasis. \synthesize is \emph{not} a drop-in replacement for an engineer. Its goal is to reduce the marginal cost of adding a feature from weeks to hours, while keeping a human reviewer at the points where judgment matters: initial specification fields, stage failures, and ingestion. \synthesize is also \emph{not} a monolithic \gls{llm} prompt: each stage is owned by a distinct agent or skill, can be inspected in isolation, and can be replaced (e.g., swapping the \texttt{KPM-Implementer} for a different codebase) without disrupting the rest of the pipeline. Two end-to-end case studies for \synthesize (an E2SM-\gls{kpm} measurement and a cross-stack Conditional Handover with E2SM-RC xApp) are presented as qualitative results in Sec.~\ref{sec:res:case:kpm} and Sec.~\ref{sec:res:case:cho}. Other capabilities are implemented according to the same logic, keeping the separation between steps (as shown in Fig.~\ref{fig:design:genesis_capabilities}) and human reviews to avoid hallucinations. 

\section{Case Study 1: Measuring and Reporting KPMs}
\label{sec:res:case:kpm}

We demonstrate \genesis on three use cases. For each one, we declare success when the pipeline runs to completion under a single natural-language prompt with the human reviewer accepting at the documented gates, the synthesized code compiles and passes the milestone sequence on every targeted tier, and the \analyzer-emitted verdict matches the acceptance criteria distilled from the specifications. In the following three sections, we discuss each use case, including results that validate the \emph{output} of the \genesis pipeline on the use case. The \genesis pipeline itself is evaluated in Sec.~\ref{sec:results}.

For the first use case, we prompt the system to implement the \texttt{RRC.ConnMean} \gls{kpm}, including its computation in the protocol stack and exposure over the O-RAN E2 interface. It is the simplest of the three \genesis use cases in this paper: two specifications (one \gls{3gpp}, one \gls{oran}), one \gls{oai} subsystem (the measurement collector), leading to an O-RAN control-plane addition in which the \gls{gnb} reports a new value but nothing reaches back into the radio stack. 

% \textcolor{blue}{
\texttt{RRC.ConnMean} measures the mean number of users in \gls{rrc}~CONNECTED mode (i.e., users with an active control-plane association to the cell) per \gls{nr} cell, averaged over a configurable reporting period. Its definition spans two specifications: \gls{3gpp}~\gls{ts}~28.552, clause~5.1.1.4.1, which defines the measurement semantics (description, collection method, data type, measured object, applicable node types), and O-RAN.WG3.E2SM-KPM~\cite{orane2smkpm}, which defines the reporting envelope (report style and action/indication formats) used to expose it over the E2 interface. Both are summarized in Table~\ref{tab:rrcconnmean}. The user-side invocation is a single natural-language prompt to the orchestrator (Listing~\ref{lst:prompt-rrcconnmean}).

\begin{table}[t]
\centering
\caption{Summary of the \texttt{RRC.ConnMean} measurement used in
Case~A.}
\label{tab:rrcconnmean}
\footnotesize
\begin{tabular}{@{}lp{4.6cm}@{}}
\toprule
\textbf{Property} & \textbf{Value} \\
\midrule
Description & Mean number of users in \gls{rrc}~CONNECTED mode per
\gls{nr} cell during the granularity period \\
\gls{3gpp} reference & \gls{ts}~28.552, clause~5.1.1.4.1%
~\cite{ts28552} \\
\gls{oran} reference & O-RAN.WG3.E2SM-KPM~\cite{orane2smkpm} \\
Collection method & SI (sampling + arithmetic mean) \\
Data type & Single integer \\
Measured object & \texttt{NRCellCU} \\
Applicable node types & \texttt{ngran\_gNB},
\texttt{ngran\_gNB\_CU} \\
\gls{kpm} report style~\cite{feraudo2026xdevsm} & Style~1 (E2 Node Measurement) \\
Action / Indication formats~\cite{feraudo2026xdevsm} & Format~1 / Format~1 \\
\bottomrule
\end{tabular}
\end{table}

\begin{figure}[t]
\begin{lstlisting}[caption={End-to-end \synthesize invocation for
Case~1.},
label={lst:prompt-rrcconnmean}]
Implement the KPM measurement RRC.ConnMean
using synthesize capability. You can find more
information in TS 28.552 and O-RAN.WG3.E2SM-KPM.
\end{lstlisting}
\end{figure}

\paragraph*{Pipeline trace}

\begin{figure}[!t]
  \centering
\includegraphics[width=0.98\columnwidth]{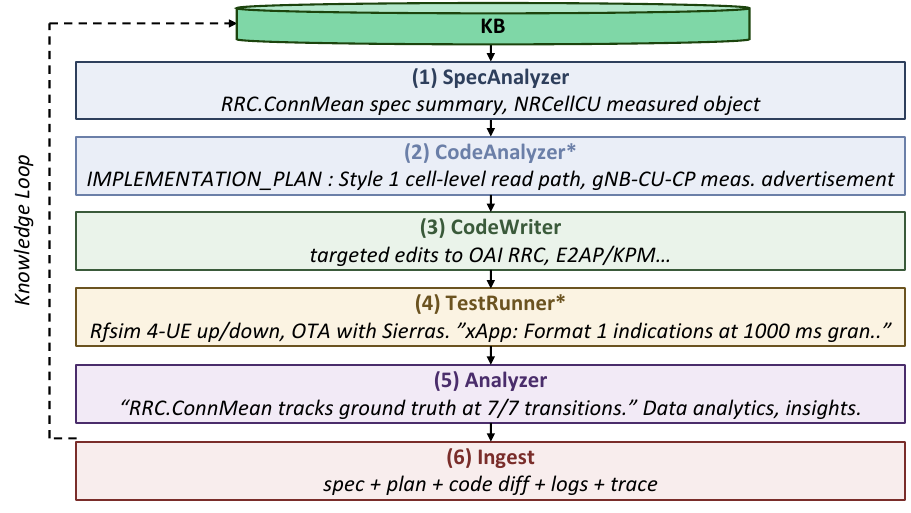}
\caption{Per-stage \synthesize pipeline trace for Case Study 1 and the RRC.ConnMean-specific artifact each produced. Stages marked by an asterisk require a human approval.}
\label{fig:design:kpm_synthesize_trace}
\end{figure}

Figure~\ref{fig:design:kpm_synthesize_trace} summarizes the per-stage trace for this case. \specanalyzer writes \texttt{/specs/RRC\_ConnMean.md} grounded against \gls{ts}~28.552. \codeanalyzer dispatches \texttt{KPM-Implementer}, which classifies the target as a \emph{snapshot value reported per cell at each tick} (the simplest of the four canonical measurement shapes the implementer recognizes), and \texttt{KPM-E2Advertiser}, which extends the list of measurements the \gls{gnb} advertises when it first joins the \gls{ric}. \codeanalyzer emits an \texttt{IMPLEMENTATION\_PLAN}, and the session pauses for user approval. On approval, \codewriter applies the plan paired with the \texttt{RAN} specialist for the compile loop. Based on our experiments (see Sec.~\ref{sec:results}), the resulting patch modifies three files in the \gls{oai} measurement subsystem. \testrunner then runs the simulation tier and replays the same flow at channel emulation and \ota using the configuration contract of Sec.~\ref{sec:testbed:staged}. \analyzer parses the bring-up milestones in the \gls{gnb} log, the stream of values reported to the \gls{ric}, and the delivery-latency distribution, and emits a pass verdict. \ingest stages the artifacts into \synapse.

\paragraph*{Outcome} 

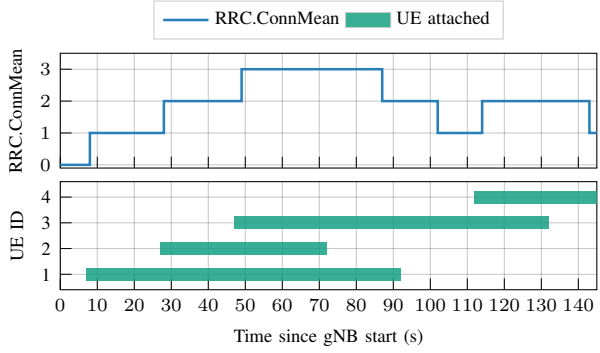
\begin{figure}[!t]
    \centering
    \begin{tikzpicture}

\definecolor{chocolate23012634}{RGB}{230,126,34}
\definecolor{darkcyan22160133}{RGB}{22,160,133}
\definecolor{darkgray176}{RGB}{176,176,176}
\definecolor{darkorchid14268173}{RGB}{142,68,173}
\definecolor{firebrick1925743}{RGB}{192,57,43}
\definecolor{lightgray204}{RGB}{204,204,204}
\definecolor{steelblue41128185}{RGB}{41,128,185}

\begin{groupplot}[
  group style={
    group size=1 by 2,
    vertical sep=0.18cm,
    x descriptions at=edge bottom,
  },
  width=0.98\columnwidth,
  label style={font=\scriptsize},
  tick label style={font=\scriptsize,/pgf/number format/fixed},
  legend cell align={left},
  legend style={
    at={(0.5,1.12)},
    anchor=south,
    font=\scriptsize,
    draw=lightgray204,
    fill=white,
    fill opacity=0.96,
    draw opacity=1,
    text opacity=1,
  },
  legend columns=2,
  scaled ticks=false,
  tick align=inside,
  tick pos=left,
  xmin=0,
  xmax=145,
  xtick distance=10,
  xtick style={color=black},
  ytick style={color=black},
  x grid style={black, opacity=0.22},
  y grid style={black, opacity=0.22},
  xmajorgrids,
  ymajorgrids,
]

\nextgroupplot[
  height=0.35\columnwidth,
  ylabel={RRC.ConnMean},
  ymin=-0.1,
  ymax=3.5,
  ytick distance=1,
]

\addplot [
  const plot,
  line width=0.96pt,
  steelblue41128185,
]
table {%
0    0
8    1
28   2
49   3
87   2
102  1
114  2
143  1
162  1
};
\addlegendentry{RRC.ConnMean}

\addlegendimage{
  area legend,
  fill=darkcyan22160133,
  draw=none,
  fill opacity=0.85
}
\addlegendentry{UE attached}

\nextgroupplot[
  height=0.34\columnwidth,
  xlabel={Time since gNB start (s)},
  ylabel={UE ID},
  ymin=0.4,
  ymax=4.6,
  ytick={1,2,3,4},
  yticklabels={1,2,3,4},
]

\draw[
  fill=darkcyan22160133,
  draw=none,
  fill opacity=0.85
]
(axis cs:7,0.75) rectangle (axis cs:92,1.25);

\draw[
  fill=darkcyan22160133,
  draw=none,
  fill opacity=0.85
]
(axis cs:27,1.75) rectangle (axis cs:72,2.25);

\draw[
  fill=darkcyan22160133,
  draw=none,
  fill opacity=0.85
]
(axis cs:47,2.75) rectangle (axis cs:132,3.25);

\draw[
  fill=darkcyan22160133,
  draw=none,
  fill opacity=0.85
]
(axis cs:112,3.75) rectangle (axis cs:147,4.25);

\end{groupplot}
\end{tikzpicture}
    \caption{Time series plot of RRC.ConnMean. The trace climbs 0-1-2-3 as UE 1–UE 3 attach, falls back to 1 as UE 2 and UE 1 detach, recovers to 2 when UE 4 attaches, and falls again to 1 when UE 3 detaches and matches the hand-computed ground truth (the number of active \texttt{nr-uesoftmodem} processes shown in the bottom panel) at every one of the seven transitions within one granularity period.}
    \label{fig:kpm_connmean_trace}
\end{figure}

The new measurement is advertised by the \gls{gnb} and observed in the reporting stream on all three validation tiers. The numerical value tracks a hand-computed ground truth within sampling tolerance as shown in Fig.~\ref{fig:kpm_connmean_trace}. The per-stage wall-clock breakdown and token cost appear in Sec.~\ref{sec:results}, where we evaluate \genesis itself.

\section{Case Study 2: Conditional Handover with Optimization xApp}
\label{sec:res:case:cho}

\begin{figure}[!t]
  \centering
  \includegraphics[width=0.98\columnwidth]{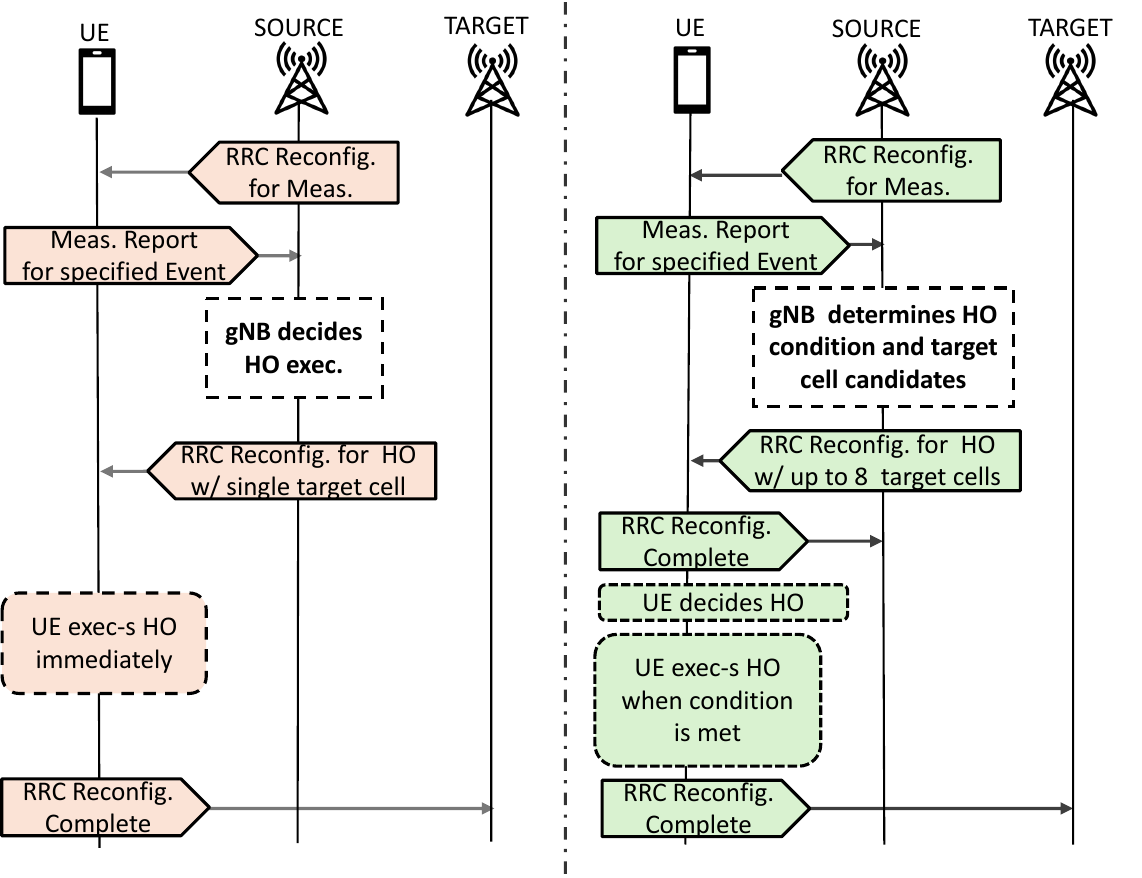}
  \caption{Traditional Handover vs. Conditional Handover in 5G NR~\cite{ts38331,ts38300}. In the traditional procedure, the network commits to a single target cell and the \gls{ue} executes the reconfiguration immediately. In \gls{cho}, the network pre-arms several candidate cells and defers the execution to the \gls{ue}.}
  \label{fig:design:ho_vs_cho}
\end{figure}

The second use case involves a more complex procedure. The implementation, testing, and optimization of \emph{Conditional Handover with a closed-loop E2SM-RC xApp} draws on four 3GPP/O-RAN specifications, touches several \gls{oai} subsystems, and terminates in an xApp (an application running in the O-RAN Near-RT RIC and controlling the stack through E2) that writes back into the \gls{gnb} to steer handover behavior at runtime. 

\gls{cho}, introduced in 3GPP Release~16~\cite{ts38331}, is a mobility procedure in which the network prepares a handover command in advance, and the \gls{ue} executes it only when its measurements satisfy a triggering condition. We illustrate this procedure in Fig.~\ref{fig:design:ho_vs_cho}, where we also compare it with the classical procedure. In the latter, the network decides when the handover happens and the \gls{ue} follows this decision. The main advantage of \gls{cho} is that the reduced overhead makes it workable in situations where regular handover fails, such as high-speed situations. Here, quick decisions from the UE are more successful compared to a lengthy exchange with the network.

\begin{figure}[!t]
  \centering
\includegraphics[width=0.98\columnwidth]{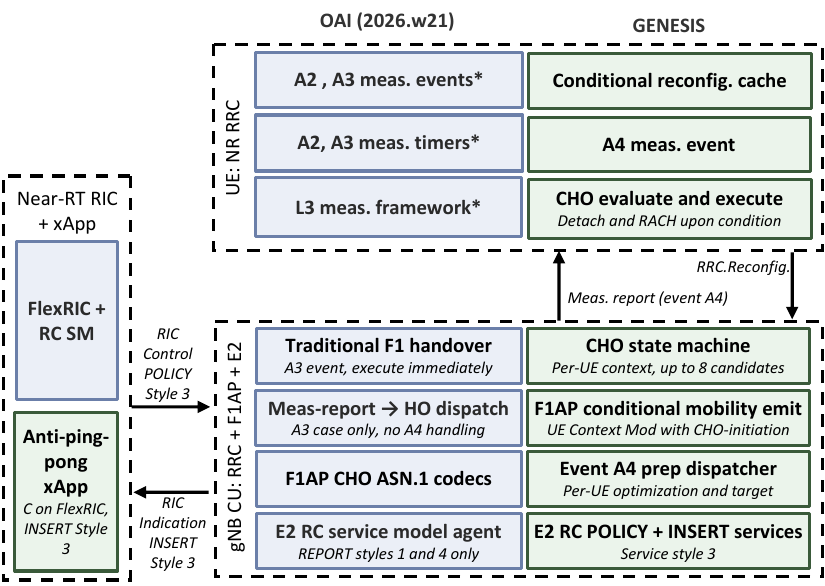}
  \caption{Logical architecture of the end-to-end \gls{cho} use case. Solid boxes (left) mark baseline OAI 2026.w21 components. Dashed boxes (right) mark modules synthesized by \genesis. Components marked by an asterisk are contributed by OAI MR !3879 and not part of the 2026.w21 baseline. The additions span three layers: a \gls{ue}-side conditional- reconfiguration cache with an A4 evaluator, a \gls{cu}-side CHO state machine that emits the F1AP CHO ASN.1 sequence and exposes new E2SM-RC POLICY/INSERT Style~3 endpoints, and a FlexRIC-hosted anti-ping-pong xApp that closes the loop over E2.}
  \label{fig:design:genesis_vs_oai}
\end{figure}

\begin{table}[t]
\centering
\caption{Summary of the Specifications for \gls{cho}}
\label{tab:cho-e2sm-rc}
\footnotesize
\begin{tabular}{@{}lp{5.0cm}@{}}
\toprule
\textbf{Property} & \textbf{Value} \\
\midrule
Description & A mobility procedure where the handover command
is prepared in advance by the network but executed by the \gls{ue} only when specific radio conditions are met \\
\gls{3gpp} references & \gls{ts}~38.300\cite{ts38300};
\gls{ts}~38.331 \cite{ts38331};
\gls{ts}~38.473 \cite{ts38473} \\
\gls{oran} reference & O-RAN.WG3.E2SM-RC-R003-v03.00
~\cite{oranrc} \\
Feature class & \gls{rrc} + F1AP + E2SM-RC + \gls{ue} \\
Trigger / execute & Event~A3/A4/A5. \\
E2 service style & \texttt{insert} Style~3 Indication~ID~2 for
mobility events; \texttt{policy} Style~3 Action~2 for CHO control \\
\bottomrule
\end{tabular}
\end{table}

To realize \gls{cho} end-to-end, \synthesize must read across four specifications, which we sum up in Table~\ref{tab:cho-e2sm-rc} (3GPP's \gls{rrc}, F1, and overall NR-NG-RAN descriptions, plus O-RAN's E2SM-RC service model that defines the control surface exposed to xApps), and implement code across several \gls{oai} subsystems (the \gls{cu}, the \gls{du}, the \gls{ue}-side evaluator), as illustrated in Fig.~\ref{fig:design:genesis_vs_oai}. In this use case, we also instruct \genesis to exercise \gls{cho} from an xApp which suppresses ping-pong handovers. This ping-pong phenomenon arises when a \gls{ue} is in between two cells, and alternatively crosses \gls{kpi} thresholds for triggering a handover, which makes it perform multiple handovers between the cells in a row, yielding to poor performance. The xApp itself is based on the OAI FlexRIC implementation, showing how \genesis can work across different code bases. This whole process is triggered by a single natural-language prompt (Listing~\ref{lst:prompt-cho}), augmented only by pointers to the specifications.

\begin{figure}[t]
\begin{lstlisting}[caption={End-to-end \synthesize invocation for
Case 2.},
label={lst:prompt-cho}]
Implement Conditional Handover over F1 end-to-end
together with a closed-loop E2SM-RC anti-ping-pong
xApp. Design relevant simulation experiments to 
test yourself and handoff to me for the OTA test.
You can find more information in TS 38.300, 
TS 38.331, TS 38.473, and O-RAN.WG3.E2SM-RC-R003.
\end{lstlisting}
\end{figure}

\paragraph*{What \genesis produced}

\begin{figure}[!t]
  \centering
\includegraphics[width=0.98\columnwidth]{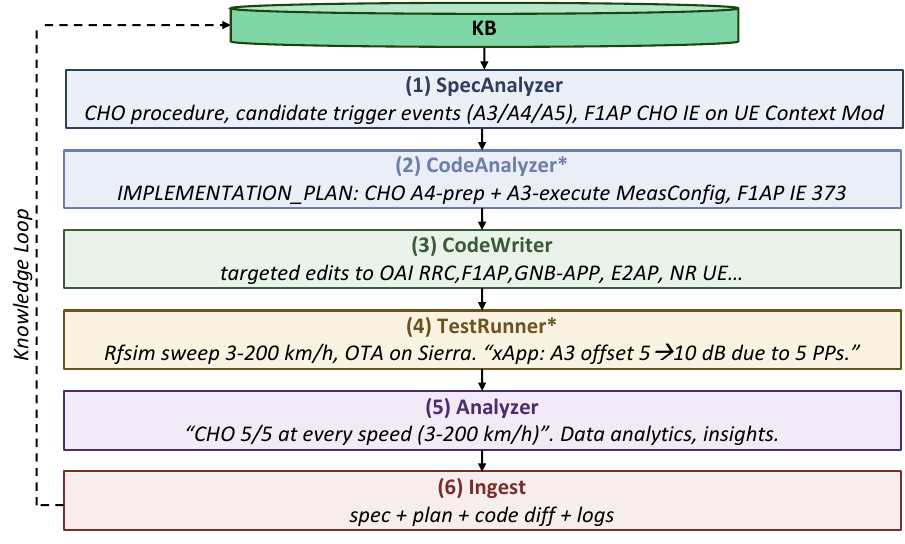}
  \caption{Per-stage \synthesize pipeline trace for Case Study 2 and the CHO-specific artifact each produced. Stages marked by an asterisk require a human approval.}
  \label{fig:design:cho_synthesize_trace}
\end{figure}

Figure~\ref{fig:design:cho_synthesize_trace} traces the six \synthesize stages and the CHO-specific artifact each produced. The rest of this paragraph walks through them in order. \specanalyzer queries each specification in turn and writes \texttt{/specs/cho-e2sm-rc.md}, capturing four artifacts: a plain description of the conditional-handover procedure, the trigger condition the \gls{ue} must evaluate (\genesis always queries the human reviewer for the trigger to use, and we always chose a signal-strength inequality known in 3GPP as ``Event A4'' in all runs), the message shape that carries the prepared command between \gls{gnb} units, and the control surface the xApp will use to push commands into the \gls{gnb}. \texttt{DevOps} then brings up the \gls{gnb} \gls{cu} pod and builds the relevant container image. \codeanalyzer locates the insertion surface in the \gls{oai} source (the existing non-conditional handover code, the data structures the \gls{gnb} uses to advertise its capabilities to the \gls{ric}, the \gls{ue}-side measurement evaluator, and the pre-compiled codec for the inter-unit messages) and emits an \texttt{IMPLEMENTATION\_PLAN} that designs the architecture shown in Fig.~\ref{fig:design:genesis_vs_oai}.

The architecture identified by \genesis and synthesized contribution spans four layers: (i) a \gls{cu}-side CHO module on the order of a dozen new functions and supporting data structures that prepares, arms, and emits the conditional reconfiguration over F1AP; (ii) a UE-side evaluator that fires once the cached conditional reconfiguration's trigger is met; (iii) two new E2SM-RC service points on the gNB E2 agent (an INSERT-style indication on every handover event and a POLICY-style control accepting an updated A3 offset); and (iv) the xApp itself, hosted on the open-source FlexRIC runtime. The xApp subscribes to the INSERT endpoint; on each handover event it appends to a per-\gls{ue} ring buffer and applies a pair-wise ping-pong rule. The xApp flags a ping-pong whenever two consecutive handovers form an A-B-A pattern within $T_\text{pp}=10$\,s. On detection, the xApp pushes a control message back to the \gls{gnb} that raises the signal-strength margin required to trigger another handover (A3 offset) between the affected cells by $+1$\,dB, damping the oscillation.

\begin{figure}[!t]
    \centering
    \input{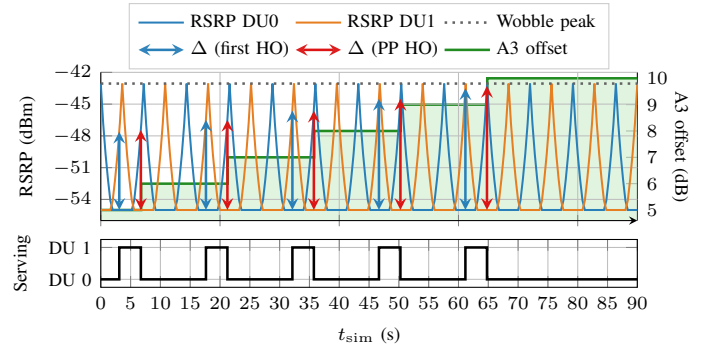}
    \vspace{-.6cm}
    \caption{xApp anti-ping-pong loop. UE wobbles between $x=4$\,m and $x=16$\,m on a 20\,m straight line at 12\,km/h. The A3 offset rises by 1\,dB per detected ping-pong. Once the offset reaches 10\,dB, the corresponding threshold %($\text{offset} + 2\,\text{dB}$ hys $=12$\,dB)
    exceeds the wobble's maximum gap and the UE remains on its last serving cell. Blue / red arrows distinguish the first HO of each mirror pair from the back HO that completes the ping-pong.}
    \label{fig:cho-xapp}
\end{figure}

\paragraph*{Closed-loop validation in simulation}

\testrunner first stress-tests the xApp in a two-cell \rfsim deployment: two \glspl{du} placed 20\,m apart on a 1-D corridor, with the simulated \gls{ue} wobbling between $x=4$\,m and $x=16$\,m at 12\,km/h. Figure~\ref{fig:cho-xapp} depicts the trial. Every crossing of the cell boundary fires a handover and the return leg of the wobble produces the mirror handover an instant later. The xApp detects five such pairs in succession and, after each, pushes a $+1$\,dB margin-bumping update back to the \gls{gnb}, walking the A3 offset from 5\,dB up to 10\,dB. With the 2\,dB hysteresis added on top, the effective trigger threshold reaches 12\,dB. The next boundary crossing no longer satisfies the A3 entry condition and the \gls{ue} stays on its last serving cell.
% } 
The trial confirms that the synthesized control loop closes correctly: events from the \gls{gnb} are decoded by the xApp, ping-pongs are classified accurately, and control messages land back at the \gls{gnb} in time to influence the next decision. 

\begin{figure}[!t]
    \centering
    \begin{tikzpicture}

\definecolor{chocolate23012634}{RGB}{230,126,34}
\definecolor{darkcyan22160133}{RGB}{22,160,133}
\definecolor{darkgray176}{RGB}{176,176,176}
\definecolor{darkorchid14268173}{RGB}{142,68,173}
\definecolor{firebrick1925743}{RGB}{192,57,43}
\definecolor{lightgray204}{RGB}{204,204,204}
\definecolor{steelblue41128185}{RGB}{41,128,185}

\begin{axis}[
width=0.98\columnwidth,
height=0.4\columnwidth,
font=\scriptsize,
label style={font=\scriptsize},
tick label style={font=\scriptsize,/pgf/number format/fixed},
legend cell align={left},
legend style={
  at={(0.5,1.02)},
  anchor=south,
  font=\scriptsize,
  draw=lightgray204,
  fill=white,
  fill opacity=0.96,
  draw opacity=1,
  text opacity=1
},
legend columns=2,
scaled ticks=false,
tick align=inside,
tick pos=left,
ybar,
ymajorgrids,
bar width=10pt,
symbolic x coords={3,30,60,120,200},
xtick=data,
x grid style={darkgray176},
xlabel={Speed (km/h)},
xtick style={color=black},
y grid style={darkgray176},
ylabel={HO Succ (\%)},
ymin=0,
ymax=105,
ytick={0,20,40,60,80,100},
ytick style={color=black},
enlarge x limits=0.1,
]
\addplot+[
  fill=firebrick1925743!65,
  draw=firebrick1925743!90!black,
  nodes near coords style={/tikz/text=black}
] coordinates {
  (3,100)
  (30,100)
  (60,100)
  (120,100)
  (200,100)
};
\addlegendentry{CHO}

\addplot+[
  fill=darkcyan22160133!65,
  draw=darkcyan22160133!90!black,
  nodes near coords style={/tikz/text=black}
] coordinates {
  (3,100)
  (30,100)
  (60,80)
  (120,40)
  (200,0)
};
\addlegendentry{HO}

\end{axis}
\end{tikzpicture}
\label{fig:cho_performance}
    \caption{Handover success rate as a function of \gls{ue} speed.
    For each (mode, speed) pair, five independent experiments were
    conducted. HO is considered successful when the \gls{ue}
    completes RACH on the target cell and the \gls{cu} logs handover
    completion.}
    \label{fig:cho-performance}
\end{figure}
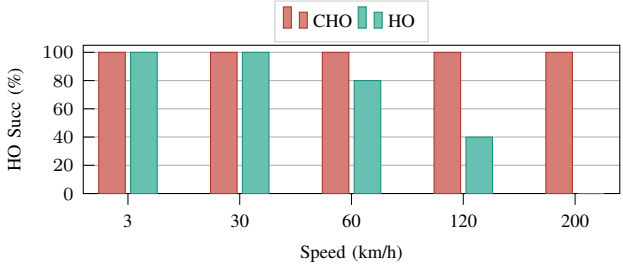

\paragraph*{High-speed two-cell deployment}
To evaluate the benefits of \gls{cho}, \testrunner next builds scenarios with increasing \gls{ue} speeds: two cells on the same \gls{nr} frequency (band~78), one \gls{cu}, two \glspl{du}, and one soft\gls{ue}, all connected over \rfsim. The DU cell sites are 200 m apart. The \gls{ue} moves from one to the other in a loop, and completes one handover per transit leg. The scenario is replayed at five speeds, from 3\,km/h (pedestrian) to 200\,km/h (high-speed train). \genesis then compares the handover success rate for each condition and handover style by running each experiment five times. Figure~\ref{fig:cho-performance} reports the handover success rate as a function of \gls{ue} speed. We observe that the \gls{cho} implemented by \genesis successfully improves the reliability of the \gls{ran} for high speed scenarios: in all conditions, \gls{cho} always succeeds, while at 60 km/h, 20\% of classic handovers fail. This rises to 60\% of failure at 120 km/h and 100\% at 200 km/h.

\paragraph*{Over-the-air validation}

\begin{figure}[t]
    \centering
\includegraphics[width=0.9\columnwidth]{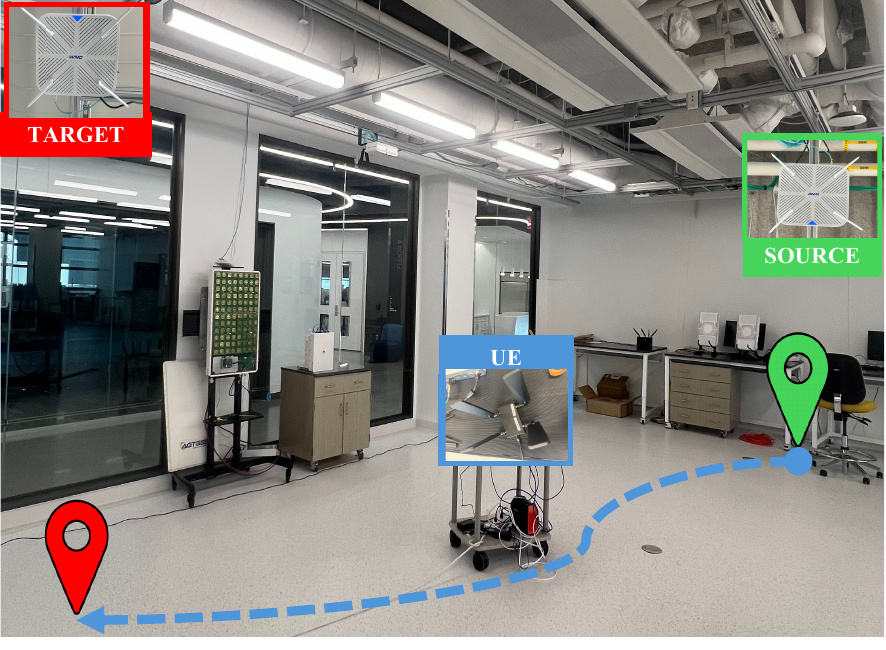}
    \caption{Over-the-air \gls{cho} testbed setup. The source DU  and target DU are deployed at opposite corners of the lab. A Sierra Wireless EM9191 \gls{ue} on a mobile cart is walked along the dashed path from the source's coverage zone into the target's.}
    \label{fig:cho-sierra-room}
\end{figure}

\begin{figure}[!t]
    \centering
    \begin{tikzpicture}

\definecolor{chocolate23012634}{RGB}{230,126,34}
\definecolor{darkcyan22160133}{RGB}{22,160,133}
\definecolor{darkgray176}{RGB}{176,176,176}
\definecolor{darkorchid14268173}{RGB}{142,68,173}
\definecolor{firebrick1925743}{RGB}{192,57,43}
\definecolor{lightgray204}{RGB}{204,204,204}
\definecolor{steelblue41128185}{RGB}{41,128,185}

\begin{axis}[
width=\columnwidth,
height=0.4\columnwidth,
font=\scriptsize,
label style={font=\scriptsize},
tick label style={font=\scriptsize,/pgf/number format/fixed},
legend cell align={left},
legend style={
  at={(0.45,1.05)},
  anchor=south,
  font=\scriptsize,
  draw=lightgray204,
  fill=white,
  fill opacity=0.96,
  draw opacity=1,
  text opacity=1
},
legend columns=4,
scaled ticks=false,
tick align=inside,
tick pos=left,
x grid style={darkgray176},
xlabel={Time since UE attach to source DU (s)},
xmin=0,
xmax=30,
xtick style={color=black},
y grid style={darkgray176},
xtick distance = 3,
ylabel={RSRP (dBm)},
ytick distance=5,
ymajorgrids,
xmajorgrids,
ymin=-100,
ymax=-65,
ytick style={color=black},
ylabel shift=-5pt
]

\addplot [
  line width=0.6pt,
  chocolate23012634
]
table {%
0 -75
30 -75
};
\addlegendentry{A4 threshold}

\addplot [
  line width=1.3pt,
  chocolate23012634,
  opacity=0.85,
  dotted
]
table {%
14.79 -110
14.79 -60
};
\addlegendentry{A4 fires}

\addplot [
  line width=1.3pt,
  darkorchid14268173,
  opacity=0.85,
  dotted
]
table {%
16 -110
16 -60
};
\addlegendentry{A3 fires}

\addplot [
  line width=1.3pt,
  darkcyan22160133,
  opacity=0.85,
  dotted
]
table {%
18.61 -110
18.61 -60
};
\addlegendentry{CHO execute}

\addplot [
  semithick,
  chocolate23012634,
  opacity=0.55,
  dashed,
  mark=square*,
  mark size=1.5,
  mark options={solid}
]
table {%
14.7 -95
14.79 -74
16 -72.5
17 -71.5
18 -71
18.61 -70
};
\addlegendentry{Candidate DU}

\addplot [
  darkorchid14268173,
  line width=1.3pt,
  <->,
  forget plot=false
]
table {%
16 -86
16 -72.5
};
\addlegendentry{$\Delta \geq $A3 offset}

\addplot [
  line width=0.96pt,
  firebrick1925743,
  mark=*,
  mark size=1.5,
  mark options={solid}
]
table {%
0.29 -86
1.57 -86
2.85 -86
4.13 -83
5.41 -79
6.69 -78
7.97 -79
9.25 -79
10.53 -80
11.81 -81
13.09 -81
14.37 -79
15.65 -85
16.93 -86
18.21 -86
};
\addlegendentry{Source DU}

\addplot [
  line width=0.96pt,
  darkcyan22160133,
  mark=triangle*,
  mark size=1.5,
  mark options={solid}
]
table {%
18.6 -70
19.88 -70
21.16 -71
22.44 -71
23.72 -71
25 -71
26.28 -71
27.56 -71
28.84 -72
29.74 -72
};
\addlegendentry{Target DU}

\end{axis}

\end{tikzpicture}
    \vspace{-.6cm}
    \caption{Over-the-air \gls{cho} result captured during one walk of the testbed in Fig.~\ref{fig:cho-sierra-room}. The \gls{cu} logs per-link RSRP for the serving cell (source DU2, PCI~1, red) and the candidate (target DU1, PCI~0, green). Three events mark the conditional-handover chain: A4 fires ($t\!\approx\!15$\,s) when the candidate crosses the A4 threshold and the network arms the conditional reconfiguration; A3 fires ($t\!\approx\!17$\,s) when the candidate exceeds the source by the A3 offset so the \gls{ue}'s local trigger is met; and CHO execute ($t\!\approx\!18$\,s) when the \gls{ue} autonomously fires the prepared handover.}
    \label{fig:cho-rsrp-timeline}
\end{figure}
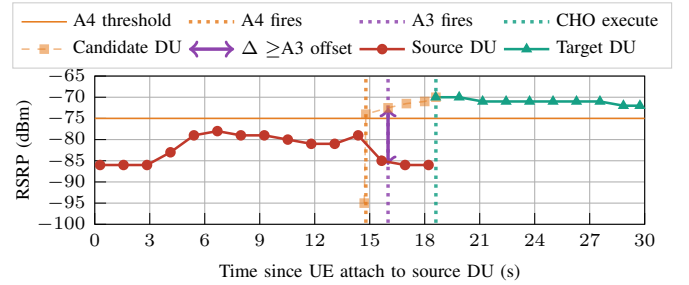

To rule out artefacts specific to \rfsim, the same two-cell band-78 deployment was brought up on real radio units on the \genesis testbed and validated \ota with a commercial Sierra Wireless EM9191 modem. Figure~\ref{fig:cho-sierra-room} shows the lab setup. Unlike the \rfsim trials, this experiment involves a human in the loop, with \genesis prepping and monitoring the experiment and prompting the human reviewer to move the UE between the two cells' coverage zones. Figure~\ref{fig:cho-rsrp-timeline} shows the resulting per-link received-signal-strength timeline collected from the \gls{cu}'s measurement reports during one such walk: the \gls{ue} attaches to DU2 (PCI~1), the prepared-but-deferred handover toward DU1 (PCI~0) is held until the cached trigger condition is satisfied, and the \gls{ue} executes the transition on its own. This validation against a COTS \gls{ue} confirms that \genesis implemented \gls{cho} in a spec-compliant way. 

\section{Case Study 3: Scheduler Discovery}
\label{sec:res:case:mac}

\begin{figure}
    \centering
    \includegraphics[width=\linewidth]{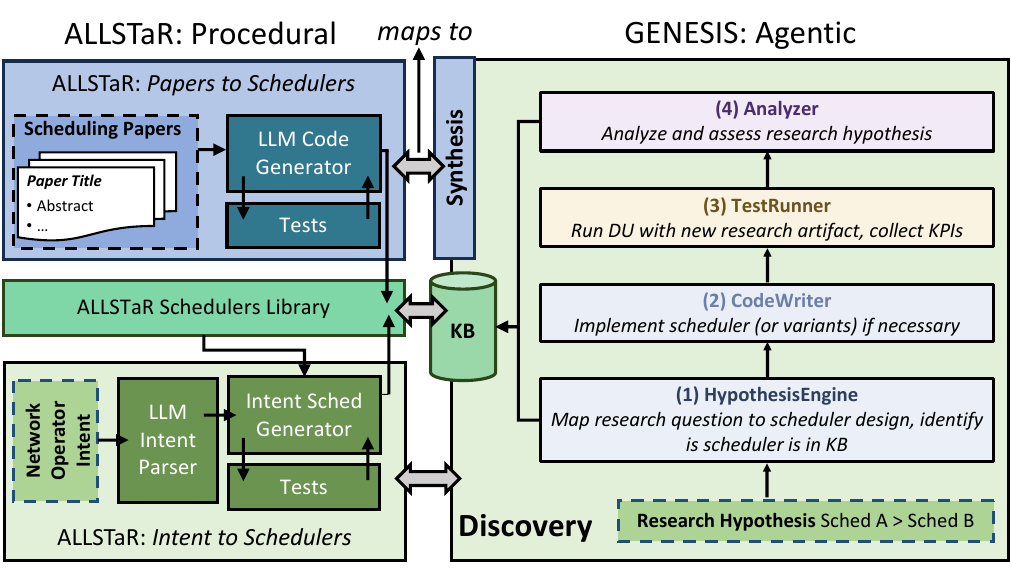}
    \caption{Mapping of ALLSTaR scheduling synthesis and discovery capabilities into the \genesis agentic discovery pipeline.}
    \label{fig:allstar}
\end{figure}

The two preceding case studies exercise \genesis end-to-end on a fresh intent. The third looks at the framework from a different angle. Of the six capabilities introduced in Sec.~\ref{sec:motivation}, this paper has shown \synthesize, \testcap, and \harden running natively. The remaining three (\discover, \optimize, \secureit) are part of the same architectural design but have not yet been driven end-to-end inside \genesis. For \discover, however, we have a close empirical reference point in our prior work \textsc{ALLSTaR}~\cite{allstar}, which built a problem-specific, procedural LLM-based pipeline for MAC scheduler synthesis using LLM agents on the same \xfiveg testbed \genesis uses. This section uses \textsc{ALLSTaR} to do two things: trace how its two procedural pipelines map onto \genesis's agentic \synthesize and \discover capabilities (Fig.~\ref{fig:allstar}), and identify the manual steps in \textsc{ALLSTaR} that \genesis could now automate.

\paragraph*{\textbf{\textsc{ALLSTaR} in brief}}
MAC scheduling is a well-studied topic in academia, with new algorithms published every year for different intents: delay-awareness, bursty traffic, fairness, slicing, etc. Each work typically benchmarks in simulation, with its own unique set of assumptions and input metrics, which makes extracting the algorithm, re-implementing it inside a production MAC stack, adapting it to the \glspl{kpi} and controls that stack exposes, and validating it on real radios a slow and cumbersome process. \textsc{ALLSTaR} closes this gap by ingesting 18 scheduler papers, and, using \gls{llm} pipelines, generating working scheduler code to be tested inside \gls{oai} on the \xfiveg testbed (top left, Fig.~\ref{fig:allstar}). The schedulers are then assembled as a library, which, combined with test results, can be leveraged by a second \gls{llm} pipeline. This second pipeline takes an operator intent as input and generates new schedulers adapted to the operators' intent (bottom left, Fig.~\ref{fig:allstar}). 

\paragraph*{\textbf{Mapping \textsc{ALLSTaR} onto \genesis primitives}}
\textsc{ALLSTaR} consists of two procedural pipelines (left of Fig.~\ref{fig:allstar}) that line up with two \genesis capabilities, mediated by the same scheduling element library that here plays the role of \synapse. The per-paper ingestion pipeline (\emph{Papers to Schedulers}) is an instance of \synthesize. The \emph{Intent to Schedulers} pipeline is the one that maps onto \discover. In \genesis, this loop becomes the four-stage agentic pipeline on the right of Fig.~\ref{fig:allstar}, driven by a research hypothesis of the form ``scheduler~A outperforms scheduler~B under workload~$W$'' (i.e., the operator intent restated comparatively, or as a research question). \hypoth~(1) translates the hypothesis into a concrete scheduler design and queries \synapse to check whether an artifact that already satisfies the claim exists; if so, the run skips ahead to validation. Otherwise, \codewriter~(2) implements the missing scheduler or a variant of one already in the library by using the same code-generation skills exercised in the previous case studies. \testrunner~(3) deploys the artifact on the \xfiveg \gls{du} and collects the \glspl{kpi} the hypothesis depends on, and \analyzer~(4) compares those measurements against the hypothesis and against library baselines drawn from \synapse, then writes the run (artifact, traces, verdict) back to \synapse for future \discover or \synthesize calls. 
Unlike \textsc{ALLSTaR}'s pipeline, where the four steps are hard-wired against a fixed library schema, and specific to scheduling, in \genesis they are agent specialists driven by the same orchestrator that runs \synthesize, so each step can be skipped, retried, or substituted depending on what \synapse already contains, and applied to other domains (e.g., handover optimization, as discussed in Sec.~\ref{sec:res:case:cho}).

\section{Performance Evaluation and Comparisons}
\label{sec:results}

We now evaluate \genesis's operational overhead in terms of time and token cost, and how its capabilities compare to existing state-of-the-art agentic LLM approaches (i.e. Claude Code). All metrics reported in this section are the median over $N=5$ independent trials. As frontier-model capabilities change on a timescale of weeks, the absolute numbers reported here are best read as a snapshot of mid-2026 model behavior, with the architecture rather than the absolute numbers as the durable contribution.

\subsection{Comparison with State-of-the-Art Coding Agent}
We start by evaluating \genesis against Claude Code as ``monolithic" baseline. Compared to \genesis, the monolithic approach has access to the same set of scripts (that are called by the skills of \genesis) and testbeds. However, instead of having access to our agents, skills, hooks, and staged validation, the monolithic baseline has to directly use its built-in tool calling routines to discover and leverage them.

We evaluate both \genesis and the baseline using Claude Code with Opus 4.7 and Sonnet 4.6 models. This comparison is performed on the simplest of our use cases, e.g., RRC.ConnMean KPM implementation. A trial is classified as a failure if the \gls{llm} performs destructive actions (e.g., deleting critical OpenShift deployments, which makes it unable to keep running experiments) or fails to converge before the exhaustion of its context window. 

We report the results in Table~\ref{tab:r3-compare}. As shown in the table, \genesis significantly improves over the baseline, which is not able to perform the task with any of the models.  We also note that Opus 4.7's success rate is of 100\% against 60\% for Sonnet 4.6, while also requiring less time to run. This poses an interesting tradeoff: if wall-clock time is the prime concern, Opus brings significant benefits, however, if one is more interested in implementing a large number of features quickly, \genesis, combined with a slightly less capable but also less expensive model (e.g., Sonnet) can have its merits, as the cost of a trial normalized by the success rate comes down to $1.4 \times 17.18 = \$$24.05 for Sonnet, against $\$28.36$ for Opus.

\begin{table}[t]
\centering
\caption{Comparison of \genesis and a monolithic single-agent baseline on the same feature-implementation task.}
\label{tab:r3-compare}
\footnotesize
\setlength{\tabcolsep}{3pt}
\begin{tabular}{@{}p{0.30\columnwidth}cccc@{}}
%\begin{tabular}{@{}lcccc@{}}
\toprule
& \multicolumn{2}{c}{\textbf{\genesis}} & \multicolumn{2}{c}{\textbf{Monolithic}} \\
\cmidrule(lr){2-3} \cmidrule(lr){4-5}
\textbf{Metric} & \textbf{Opus 4.7} & \textbf{Sonnet 4.6} & \textbf{Opus 4.7} & \textbf{Sonnet 4.6} \\
\midrule
Wall-clock                & 44 min.  & 93 min.  & 78 min.  & 113 min. \\
Success rate              & 100\%    & 60\%     & 0\%      & 0\% \\
Cost per trial (USD)      & \$28.36  & \$17.18  & \$43.76  & \$18.73 \\
\bottomrule
\end{tabular}
\end{table}

\subsection{Per-stage Cost and Time Breakdown}
\label{sec:res:perstage}

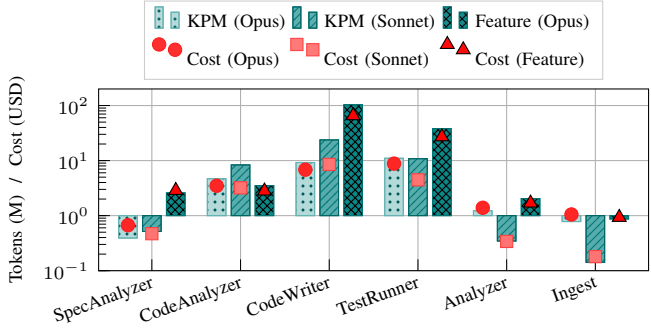
\begin{figure}[t]
\centering
\begin{tikzpicture}

\definecolor{lightgray204}{RGB}{204,204,204}
\definecolor{darkgray176}{RGB}{176,176,176}

\pgfplotsset{
  every axis/.append style={
    font=\scriptsize,
    label style={font=\scriptsize},
    tick label style={font=\scriptsize,/pgf/number format/fixed},
  },
}

\begin{axis}[
  name=tokax,
  width=\columnwidth, height=0.45\columnwidth,
  axis lines = box,
  ybar,
  ymode=log,
  bar width=7pt,
  symbolic x coords={SpecAnalyzer,CodeAnalyzer,CodeWriter,TestRunner,Analyzer,Ingest},
  xtick=data,
  x tick label style={rotate=20, anchor=north east, inner sep=1pt},
  ylabel={Tokens (M) \;/\; Cost (USD)},
  ymin=0.1, ymax=200,
  ytick={0.1,1,10,100},
  enlarge x limits=0.12,
  ymajorgrids, xmajorgrids,
  x grid style={darkgray176},
  y grid style={darkgray176},
  tick align=inside,
  tick pos=left,
  scaled ticks=false,
  xtick style={color=black},
  ytick style={color=black},
  legend cell align={left},
  legend columns=3,
  legend style={
    at={(0.5,1.05)},
    anchor=south,
    font=\scriptsize,
    draw=lightgray204,
    fill=white,
    fill opacity=0.96,
    draw opacity=1,
    text opacity=1,
  },
]

\addplot+[fill=teal!30, draw=teal!60, postaction={pattern=dots, pattern color=teal!70!black}] coordinates {
  (SpecAnalyzer,0.390869) (CodeAnalyzer,4.689142) (CodeWriter,9.224672)
  (TestRunner,11.029795) (Analyzer,1.225938)   (Ingest,0.780083)};
\addlegendentry{KPM (Opus)}

\addplot+[fill=teal!65, draw=teal!80!black, postaction={pattern=north east lines, pattern color=teal!95!black}] coordinates {
  (SpecAnalyzer,0.516230) (CodeAnalyzer,8.347230) (CodeWriter,23.752033)
  (TestRunner,10.827826) (Analyzer,0.345836)   (Ingest,0.142042)};
\addlegendentry{KPM (Sonnet)}

\addplot+[fill=teal!95, draw=teal!90!black, postaction={pattern=crosshatch, pattern color=black}] coordinates {
  (SpecAnalyzer,2.589684) (CodeAnalyzer,3.519015) (CodeWriter,103.514581)
  (TestRunner,38.082741) (Analyzer,2.042511)   (Ingest,0.868800)};
\addlegendentry{Feature (Opus)}

\addlegendimage{only marks, mark=*, mark size=2.5pt,
  mark options={fill=red!85, draw=red!90}}
\addlegendentry{Cost (Opus)}
\addlegendimage{only marks, mark=square*, mark size=2.3pt,
  mark options={fill=red!55, draw=red!75}}
\addlegendentry{Cost (Sonnet)}
\addlegendimage{only marks, mark=triangle*, mark size=3pt,
  mark options={fill=red!100!black, draw=black}}
\addlegendentry{Cost (Feature)}

\end{axis}

\begin{axis}[
  at={(tokax.south west)}, anchor=south west,
  width=\columnwidth, height=0.45\columnwidth,
  hide axis,
  ymode=log,
  symbolic x coords={SpecAnalyzer,CodeAnalyzer,CodeWriter,TestRunner,Analyzer,Ingest},
  ymin=0.1, ymax=200,
  enlarge x limits=0.12,
]
\addplot[only marks, mark=*, mark size=2.5pt,
  mark options={fill=red!85, draw=red!90, xshift=-9pt}]
  coordinates {
  (SpecAnalyzer,0.67) (CodeAnalyzer,3.49) (CodeWriter,6.84)
  (TestRunner,8.81)   (Analyzer,1.39)    (Ingest,1.05)};

\addplot[only marks, mark=square*, mark size=2.3pt,
  mark options={fill=red!55, draw=red!75}]
  coordinates {
  (SpecAnalyzer,0.47) (CodeAnalyzer,3.20) (CodeWriter,8.52)
  (TestRunner,4.47)   (Analyzer,0.34)    (Ingest,0.18)};

\addplot[only marks, mark=triangle*, mark size=3pt,
  mark options={fill=red!100!black, draw=black, xshift=9pt}]
  coordinates {
  (SpecAnalyzer,2.87) (CodeAnalyzer,2.80) (CodeWriter,65.35)
  (TestRunner,27.00)  (Analyzer,1.69)    (Ingest,0.93)};

\end{axis}

\end{tikzpicture}
\caption{Per-stage token usage and USD cost across two pipelines.
Light dotted bars: \texttt{RRC.ConnMean} \synthesize on Opus (median).
Mid striped bars: same pipeline on Sonnet (median).
Dark cross-hatched bars: CHO + E2SM-RC \synthesize on Opus.
Color encodes token category; bar saturation and fill pattern encode the run.
Cost markers (right axis) use distinct shapes per run.
Total run cost incl.\ parent session:
\textbf{\$28.36} (KPM/Opus), \textbf{\$17.18} (KPM/Sonnet),
\textbf{\$102.91} (Feature/Opus).}
\label{fig:per_stage_cost}
\end{figure}

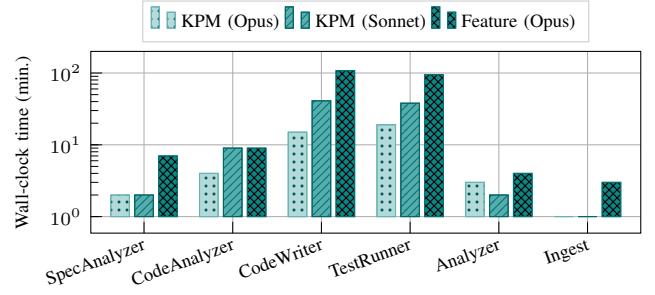
\begin{figure}[t]
\centering
\begin{tikzpicture}
\definecolor{lightgray204}{RGB}{204,204,204}
\definecolor{darkgray176}{RGB}{176,176,176}
\pgfplotsset{
  every axis/.append style={
    font=\scriptsize,
    label style={font=\scriptsize},
    tick label style={font=\scriptsize,/pgf/number format/fixed},
  },
}
\begin{axis}[
  width=\columnwidth, height=0.45\columnwidth,
  axis lines=box,
  ybar,
  ymode=log,
  bar width=7pt,
  symbolic x coords={SpecAnalyzer,CodeAnalyzer,CodeWriter,TestRunner,Analyzer,Ingest},
  xtick=data,
  x tick label style={rotate=20, anchor=north east, inner sep=1pt},
  ylabel={Wall-clock time (min.)},
  ymin=0, ymax=200,
  ytick={1,10,100},
  enlarge x limits=0.12,
  ymajorgrids, xmajorgrids,
  x grid style={darkgray176},
  y grid style={darkgray176},
  tick align=inside,
  tick pos=left,
  scaled ticks=false,
  xtick style={color=black},
  ytick style={color=black},
  legend cell align={left},
  legend columns=3,
  legend style={
    at={(0.5,1.05)},
    anchor=south,
    font=\scriptsize,
    draw=lightgray204,
    fill=white,
    fill opacity=0.96,
    draw opacity=1,
    text opacity=1,
  },
]
\addplot+[fill=teal!30, draw=teal!60, postaction={pattern=dots, pattern color=teal!70!black}] coordinates {
(SpecAnalyzer,2) (CodeAnalyzer,4)  (CodeWriter,15)
(TestRunner,19)  (Analyzer,3)      (Ingest,1)};
\addlegendentry{KPM (Opus)}
\addplot+[fill=teal!65, draw=teal!80!black, postaction={pattern=north east lines, pattern color=teal!95!black}] coordinates {
(SpecAnalyzer,2) (CodeAnalyzer,9)  (CodeWriter,41)
(TestRunner,38)  (Analyzer,2)      (Ingest,1)};
\addlegendentry{KPM (Sonnet)}
\addplot+[fill=teal!95, draw=teal!90!black, postaction={pattern=crosshatch, pattern color=black}] coordinates {
(SpecAnalyzer,7) (CodeAnalyzer,9)  (CodeWriter,108)
(TestRunner,95)  (Analyzer,4)      (Ingest,3)};
\addlegendentry{Feature (Opus)}
\end{axis}
\end{tikzpicture}
\caption{Median wall-clock time per \synthesize stage.
Light dotted bars: \texttt{RRC.ConnMean} pipeline on Opus.
Mid striped bars: same pipeline on Sonnet.
Dark cross-hatched bars: CHO + E2SM-RC pipeline on Opus.
Solo totals: \textbf{44 min.} (Opus KPM), \textbf{93 min.} (Sonnet KPM),
\textbf{226 min.} (Opus Feature).}
\label{fig:per_stage_time}
\end{figure}

We now profile the operational cost of \genesis itself. Figures~\ref{fig:per_stage_cost} and~\ref{fig:per_stage_time} report token usage with dollar cost and wall-clock time, broken down by the six \synthesize stages and color-coded by configuration: KPM/Opus, KPM/Sonnet, and CHO/Opus.

\textbf{Two stages dominate.} Across all three configurations, \codewriter and \testrunner together account for roughly two-thirds of both wall-clock time and token cost. Both are iteration-heavy: \codewriter loops against compiler output until the patch builds, and \testrunner replays the same bring-up flow at each tier of the validation continuum. \specanalyzer and \codeanalyzer are dominated by grounded retrieval against \synapse and are inexpensive by comparison. \analyzer and \ingest have the lowest expenses.

\textbf{Cache reads dominate the long-running stages.} 
Cache reads account for 94\% of all tokens consumed across the pipeline and are billed at 10\% of the base input price per token~\cite{claude-pricing}. This is the visible signature of two design decisions: the \texttt{SKILL.md} progressive-disclosure pattern reuses the same procedural body across iterations, and the specialist-isolation discipline (Sec.~\ref{sec:design:orchestration}) keeps the same persona on the same context for the duration of its sub-conversation. Without these choices the per-trial cost would be several times higher than the reported \$28.36~/~\$17.18~/~\$102.91.

\textbf{Complexity scales with feature scope, not framework overhead.} The CHO configuration costs roughly $3.6\times$ what \texttt{RRC.ConnMean} costs in dollars and $5.1\times$ in wall-clock, on the same six stages and the same orchestrator. The blow-up concentrates in \codewriter and \testrunner, which are the stages whose work is bounded by the feature, not by the framework. \specanalyzer and \codeanalyzer rise modestly (more specs to read, more insertion sites to map). \analyzer and
\ingest are essentially unchanged. The framework overhead is therefore approximately fixed and the marginal cost of harder features is paid in the stages that have to do more actual work.
% }

\subsection{Model Selection: Speed vs.\ Cost per Success}
\label{sec:res:tradeoff}

% \textcolor{blue}{
Analyzing Table~\ref{tab:r3-compare} alongside the per-stage figures surfaces a counter-intuitive operating point. Opus~4.7 is faster (44~min vs.~93~min), succeeds more often (100\% vs.~60\%), and costs more per trial (\$28.36 vs.~\$17.18). Naively, Opus dominates. But once we normalize cost by success rate the expected cost of producing one working implementation, the picture inverts. For Sonnet, the expected cost per success is $\$17.18 / 0.60 \approx \$28.63$, essentially identical to Opus's per-trial cost. Wall-clock to one expected success is $93 / 0.60 \approx 155$~min for Sonnet against $44$~min for Opus (a $3.5\times$ Opus advantage on latency, but parity on cost).

The operating point that follows: \emph{if wall-clock latency drives the deployment, Opus dominates; if the goal is to amortize a fixed budget across many features run in parallel, Sonnet is competitive on cost-per-success and trades latency for throughput}. This is the tradeoff the agent/skill/hook split is designed to enable: because each specialist's reasoning budget is bounded by its persona, swapping the underlying model is a configuration change rather than a re-implementation.

\section{Conclusion and Future Work}
\label{sec:conclusion}

In this paper, we presented \genesis, an agentic framework that compresses the cellular R\&D life-cycle into a closed, intent-driven loop grounded in real-radio observations and a persistent knowledge plane. We built \genesis on three portable primitives (agents, skills, and hooks) coordinated by an orchestrator and unified by \synapse, a hybrid-retrieval knowledge base that serves as both ground truth and artifact sink. \genesis drives a three-tier validation continuum (\rfsim, channel emulation, \ota on \xfiveg) and six capability pipelines (\synthesize, \testcap, \harden, \optimize, \discover, \secureit) spanning the full RAN R\&D life-cycle. We instantiated three end-to-end case studies, i.e., \texttt{RRC.ConnMean} as a \synthesize anchor, conditional handover with a closed-loop E2SM-RC xApp, and the ALLSTaR scheduler-discovery pipeline. For synthesis, \genesis achieved a 100\% success rate across statistically independent trials, while the off-the-shelf Claude Code baseline (Opus~4.7 and Sonnet~4.6) produced no working implementation on any attempt. 
Next steps include further development of agentic capabilities, especially around \discover and \secureit instantiations, deeper cross-capability compounding through \synapse, and closing the standards feedback loop with \genesis-driven 3GPP contributions. 

\balance
\bibliographystyle{IEEEtran}
\bibliography{references}

\end{document}